\documentclass{emulateapj}
\usepackage{natbib} 
\newcommand{\alc}[1]{{#1}}
\newcommand{\chg}[1]{#1}
\newcommand{\chga}[1]{#1}

\slugcomment{Accepted to ApJ}

\shorttitle{Spatial Clustering of RASS AGNs II --  HOD Modeling}
\shortauthors{Miyaji et al.}

\usepackage{graphicx}

\begin{document}


\title{The Spatial Clustering of ROSAT All-Sky Survey AGNs\\ 
II. Halo Occupation Distribution Modeling of the Cross Correlation Function}


\author{Takamitsu Miyaji\altaffilmark{1,2}, Mirko Krumpe\altaffilmark{2},
  Alison L. Coil\altaffilmark{2,3}, Hector Aceves\altaffilmark{1}}
\altaffiltext{1}{Instituto de Astronom\'ia, Universidad Nacional
   Aut\'onoma de M\'exico, Ensenada, Baja California, M\'exico 
   (PO Box 439027, San Diego, CA 92143-9027, USA)}
\altaffiltext{2}{University of California, San Diego, 
  Center for Astrophysics and Space Sciences, 9500 Gilman Drive, 
  La Jolla, CA 92093-0424, USA}
\altaffiltext{3}{Alfred P. Sloan Foundation Fellow}
\email{miyaji@astrosen.unam.mx}



\begin{abstract}

  This is the second paper of a series that reports on our investigation
of the clustering properties of AGNs in the {\it ROSAT} All-Sky Survey (RASS) 
through cross-correlation functions (CCFs) with Sloan Digital Sky Survey 
(SDSS) galaxies. In this paper, we apply the Halo Occupation Distribution 
(HOD) model to the CCFs between the RASS Broad-line AGNs with 
SDSS Luminous Red Galaxies (LRGs) in the redshift range $0.16<z<0.36$ that 
was calculated in paper I. {In our HOD modeling approach, we use the 
known HOD of LRGs and constrain the HOD of 
the AGNs by a model fit to the CCF. For the first time, we are able 
to go beyond quoting merely a `typical' AGN host halo mass, $M_{\rm h}$, 
and model the full distribution function of AGN host dark matter halos.} 
In addition, we are able to determine the large-scale bias and the 
mean $M_{\rm h}$ more accurately. \chg{We explore the behavior of \chga{three} 
simple HOD models. \chga{Our first model (Model A) is a truncated power-law HOD 
model in which all AGNs are satellites. With this model,} we find an upper limit 
to the slope ($\alpha$) of the AGN HOD that is far below unity.  
\chga{The other two models have a central component,  which
has a step function form, where the HOD is  constant above a minimum mass, 
without (Model B) or with (Model C) an upper mass cutoff, in addition to the 
truncated power-law satellite component, similar to the HOD that is found for galaxies. 
In these two models we find that the upper limits on $\alpha$ are still 
below unity, with $\alpha\la 0.95$ and $\alpha\la 0.84$ for Model B and C respectively. 
Our analysis suggests that the satellite AGN occupation increases slower than, or 
may even decrease with, $M_{\rm h}$ in contrast to the satellite HODs of 
luminosity-threshold samples of galaxies, which, in contrast, grow approximately as 
$\langle N_{\rm s}\rangle \propto M_{\rm h}^\alpha$ with $\alpha\approx 1$. These results are consistent with 
observations that the AGN fraction in groups and clusters decreases with richness.}}

\end{abstract}






\section{Introduction}
\label{sec:intro}

 Investigating how active galactic nuclei (AGNs) are distributed in the Universe
provides a clue as to the physical conditions in which accretion onto 
supermassive black holes (SMBH) takes place. \chg{After the formation
of galaxies in the universe, at any given time 
in the history of the Universe}, only a small fraction of galaxies show AGN activity. 
It is known that almost all galaxies with a spheroidal component contain 
a SMBH, and that the BH mass is closely related to the mass or the velocity 
dispersion of the spheroidal component 
(e.g., \citealp{ferrarese_merritt00,gebhardt00,marconi03,haering04,shankar09} for review). 
This means that most galaxies have likely had one or more brief AGN periods, during 
which the central SMBH grows in such a way that the growth of the SMBH and the formation of the spheroidal component
are tightly related. The question of when and under what physical conditions 
the accretion takes place is important in understanding not only the 
origin and evolution of SMBHs but also the origin and evolution of 
galaxies. {Many models postulate that AGN activity is merger-driven,
especially those with high luminosities (i.e., QSOs), 
\citep[e.g.,][]{wyithe02,hopkins06}. On the other hand, for 
lower-luminosity AGNs, processes internal to the galaxy such as 
galaxy disk instabilities, may be 
important \citep{kauffmann07,hasinger08}.} Different mechanisms may be responsible for 
triggering AGN activity at different redshifts or luminosities.
 
 Large-scale clustering properties of AGNs provide important clues as to 
the physical process(es) that are responsible for the SMBH accretion. 
{The large-scale clustering amplitude reflects, through the bias 
parameter $b$, } the typical mass of the dark matter halos (DMH) 
in which AGN reside \citep{st99,sheth01,tinker05}. Thus exploring the clustering 
properties of AGNs at different redshifts, luminosities, 
and AGN types to probe the masses of the DMHs that host them, provides us with important 
clues in understanding constraints on SMBH growth in a cosmological context.

 Clustering measurements of X-ray point sources have been made whenever 
new large-scale surveys have been completed, through angular correlation 
functions \citep[e.g.][]{pucceti06,miyaji07,ueda08,plionis08,ebrero09b}. When redshifts 
for a complete sample become available, three-dimensional (3D) 
correlation functions
can then be computed \citep[e.g.,][]{mullis04,yang06,gilli09,cappelluti10}.  
Redshift information provides a huge leap in clustering measurements both in 
terms of the statistical accuracy as well as in removing systematic errors 
associated with model assumptions in making Limber's de-projection 
\citep{limber54}. However, even with redshift information, small number 
statistics has limited the accuracy of correlation function (CF) measurements, 
especially through the auto-correlation function (ACF) of the AGNs themselves.  
The situation can be improved by measuring the cross-correlation function (CCF) 
of AGNs with a galaxy sample that has a much higher space density. 
In order to facilitate a CCF analysis, one needs 
an extensive galaxy redshift surveys with common sky and 
redshift coverage as the AGN redshift surveys. With recent large-scale 
survey projects, measurements of AGN clustering through CCF with galaxies 
are now emerging \citep[e.g.,][]{li06,coil07,coil09,hickox09,padmanabhan09,
mountrichas09}.

 While the {\sl ROSAT} All Sky Survey (RASS) \citep{voges99}, most of which was conducted 
during the first half year of the {\sl ROSAT} mission (mostly in 1990), 
produced a catalog of $\approx 120,000$ X-ray point sources, the availability of 
a comprehensive redshift survey of RASS-selected AGNs was limited until a 
RASS-Sloan Digital Sky Survey (SDSS) matched AGN catalog became available 
\citep{anderson03,anderson07}.  In view of this, we initiated a series of 
studies of the clustering properties of low redshift AGNs in the RASS-SDSS 
sample through the CCF analysis with SDSS galaxies.
In \citet[][hereafter paper I]{paperI}, we reported our results on our CCF 
analysis between broad-line AGNs in RASS that have been 
identified with SDSS \citep{anderson07} and the SDSS Luminous Red 
Galaxies (LRGs) \citep{eisenstein01} in the redshift range $0.16<z<0.36$.

{Through these efforts to measure the AGN ACF/CCF, the masses of the DMHs 
where AGN activity occurs are being gradually uncovered. 
Most results measuring the 3D 
correlation functions of X-ray selected AGNs indicate that the typical DMH mass 
in which these AGN reside is in the range  
$12\la \log M_{\rm h} [h^{-1}M_{\rm sun}]\la 13.5$ at  
both low ($z\la 0.4$) and high ($z\approx 1$) redshifts 
\citep[e.g., paper I; ][]{coil09,gilli09,cappelluti10}. Optically-selected QSOs, typically 
representing a high redshift, high luminosity AGN population, are associated with DMHs with 
a typical mass in the range  $12\la \log M_{\rm h} [h^{-1}M_{\rm sun}]\la 13$ 
\citep[e.g.][]{porciani04,coil07,shen07,ross09}. It appears that the luminosity 
dependence of AGN clustering may be different at different redshifts. In paper I,
we show that among broad-line AGN, high X-ray luminosity AGNs are more strongly 
clustered than lower X-ray luminosity AGN at z$\approx 0.3$. At $z\approx 1$, however, 
low luminosity X-ray selected AGNs are more strongly clustered than optical QSOs 
\citep{coil09}. This apparent difference 
may be caused by a non-monotonic luminosity dependence and/or AGN type dependence 
of biasing rather than a dependence
on redshift. What is needed is to explore AGN clustering 
across wider ranges in luminosity-redshift space to break this degeneracy.}  

 In interpreting the correlation function measurements, most previous studies 
use the large scale bias ($b$) of AGNs to infer the associated typical DMH mass using 
linear growth and linear biasing schemes.  Strictly speaking, this is only valid for the 
correlation function measurements on sufficiently large scales 
($r\ga 1-2 h^{-1}$ Mpc). Non-linear modeling through the Halo Occupation Distribution 
(HOD) framework \citep[e.g.][]{peacock00,seljak00,cooray_sheth02} is imperative to accurately 
interpret and make full use of the correlation function measurements. In 
this framework, the mean number of the sample objects in the DMH 
($\langle N \rangle$) is modeled as a function of the DMH mass ($M_{\rm h}$).
Then the two-point correlation function is modeled as the sum of the 
contributions of pairs from the same DMH (1-halo term) and those from different DMHs 
(2-halo term). This method has been used extensively to interpret galaxy correlation 
functions \citep[e.g.][hereafter Z09]{hamana04,tinker05,phleps06,zheng07,zehavi10,zheng09} 
to constrain how various galaxy samples are distributed among DMHs as well as whether 
these galaxies occupy the centers of the DMHs or are satellite galaxies 
\citep{kravtsov04,zheng05}.     

\chg{Partially due to the low number density of AGNs, there have been
  \alc{few results in the literature interpreting AGN correlation functions 
    using HOD modeling, where the small-scale clustering measurements are
  essential. \citet{padmanabhan09} discussed qualitative HOD
  constraints on their LRG-optical QSO CCF, where they argued that
  the acceptable models include those in which $>25$\% of their QSOs are
  satellites and those in which QSOs are a random subsampling of a
  luminosity-threshold sample of galaxies.  \citet{shen10} also used
  the HOD modeling approach to binary pairs of QSOs at $3\lesssim z
  \lesssim 4$ and conclude that they favor a model in which $\gtrsim
  10$\% of the QSOs are satellites.}}

 In this paper, we apply the HOD model to the CCFs of the RASS broad-line AGNs
(hereafter, simply referred to as AGNs) and 
SDSS LRGs obtained in paper I. \chg{The HOD modeling has allowed us to model the 
CCFs beyond simple power-law fits and to investigate constraints on how these AGNs are 
distributed among DMHs as a function of halo mass.}
{While almost all previous HOD modeling studies apply the method to 
ACFs, here we apply the HOD modeling approach to the CCF. With this method, we use the 
previously estimated LRG HOD to model the AGN HOD by fitting to the measured AGN-LRG CCF.    
In our analysis, we take advantage of the LRG HOD of Z09, which is 
based on the well-measured LRG ACF from SDSS \citep{zehavi05b}. This approach was taken by Z09 
in modeling LRG-$L_*$ galaxy CCFs, assuming that the LRGs are central galaxies of the DMHs 
and that the $L_*$ galaxies are satellite galaxies upon the calculation of the 1-halo term. 

In this paper, we present a comprehensive explanation 
of the  HOD analysis that can be applied to general cases, and then 
we apply the method to our AGN-LRG CCF. 

 The scope of the paper is as follows. In sect. \ref{sec:ccf_results}, 
we summarize the samples used in paper I and the basic methods of CCF 
calculations. In Sect. \ref{sec:halomodel}, we explain our basic modeling 
procedure of the LRG ACF, which is used as a template to apply the HOD 
modeling to the AGN-LRG CCF. In Sect. \ref{sec:results}, we show our results 
on the AGN HODs. Sect. \ref{sec:disc} discusses our results, including a 
comparison with the results from paper I and the astrophysical implications of our 
constraints on the AGN HOD obtained here.  
Finally, Sect. \ref{sec:concl} summarizes important consequences from our
analysis and concludes our discussion.         

Throughout the paper, all distances are measured in comoving coordinates 
and given in units of $h^{-1}$\,Mpc, where $h= H_{\rm 0}/100$\,km\,s$^{-1}$. 
We also use the symbol
$h_{70}= H_{\rm 0}/70$\,km\,s$^{-1}$ for X-ray luminosities, while $h=1$ 
is used to express 
optical absolute magnitudes, for consistency with referenced articles. 
We use a cosmology with 
$h=0.7$, $\Omega_{\rm M}=0.3$, $\Omega_{\rm \Lambda}=0.7$,
$\sigma_8=0.8$, $\Omega_{\rm baryon}=0.047$ and $n_{\rm s}=1$, which are 
consistent with the most updated WMAP cosmology as of writing this paper 
\citep{spergel03}\footnote{\url{http://lambda.gsfc.nasa.gov/product/map/current/parameters.cfm}}. 
The symbol $\log$ signifies a base-10 logarithm, while a natural logarithm
is expressed by an $\ln$.

\section{The RASS AGN-LRG CCF Measurements} 
\label{sec:ccf_results}

In paper I, we calculate the CCFs between AGNs and the LRG, as well as
the LRG ACF in the redshift range $0.16<z<0.36$. Basic properties of the sample used 
in that paper are repeated here in Table \ref{tab:samp}. The rationale 
and strategies of the sample selection are discussed in paper I in detail,
\chg{including a redshift-$L_{\rm x}$ diagram of the AGN sample 
(Fig. 1 of paper I).} Here we briefly summarize the sample selection and CF calculations below.

 We use the AGN sample from the RASS-SDSS matched AGN catalog by 
\citet{anderson03,anderson07}, which is based on the SDSS Data Release (DR) 5.
In order to make a well-defined, uniform AGN sample, we only include 
broad lines AGNs. For the reference sample, we extract LRGs from the 
SDSS Catalog Archive Server Jobs System\footnote{\url{http://casjobs.sdss.org/CasJobs/}}
using the flag ``galaxy\_red'', {which is based on} the selection criteria defined in
\citep{eisenstein01}. We create a volume-limited spectroscopic LRG
sample with $-23.2 < M_{g}^{z=0.3} < -21.2$ and $0.16 <z<0.36$.
The AGN sample is also limited to this redshift range.  In order to 
obtain the largest common geometry between the AGNs and LRGs with public 
geometry and completeness files \citep{blanton05}, we limit our sample to  
the SDSS DR4+ geometry \footnote{\url{http://sdss.physics.nyu.edu/lss/dr4plus}}.  
The selected geometry covers an area of 5468 deg$^{2}$. The AGN sample 
is further divided into high $L_{\rm X}$ and
low $L_{\rm X}$ samples. The numbers of LRGs and AGNs used in our CCF measurements are 
listed in Table \ref{tab:samp}.


\begin{deluxetable*}{ccccccc}
\tabletypesize{\normalsize}
\tablecaption{Properties of the LRG and RASS-AGN samples\label{tab:samp}}
\tablewidth{0pt}
\tablehead{
\colhead{sample} & \colhead{$z$}  & \colhead{$M_{g}^{z=0.3}$ range} & \colhead{} & \colhead{$<n_{\rm LRG}>$} & \colhead{} & \colhead{$<M_{g}^{z=0.3}>$}\\
\colhead{name} & \colhead{range}  & \colhead{[mag]} & \colhead{number} & \colhead{[$h^{3}$ Mpc$^{-3}$]} & \colhead{$<z>$} & \colhead{[mag]}}
\startdata
LRG sample            & $0.16 <z<0.36$ & $-23.2 < M_{g}^{z=0.3} < -21.2$  & 45899 & $9.6 \times 10^{-5}$ & 0.28& -21.71 \\\hline
                      &                &                                &       &                      &     & \\    
                      &                & $\log L_{\rm X}$ range\tablenotemark{a,b}&       & $<n_{\rm AGN}>$       &     & $\log <L_{\rm X}>$\tablenotemark{a,c}\\
                      & $z$-range      & range [$h_{70}^{-2}$erg s$^{-1}$]  & number& [$h^{3}$ Mpc$^{-3}$] & $<z>$& [$h_{70}^{-2}$ erg s$^{-1}$]\\\hline

\\
All RASS-AGN sample & $0.16 <z<0.36$ & $43.7\la \log L_{\rm X}\la 44.7$     &  1552 & $6.0 \times 10^{-5}$ & 0.25 & 44.17 (44.16)\\
High $L_{\rm X}$ RASS-AGN sample & $0.16 <z<0.36$ &$44.3<\log L_{\rm X} \la 44.9$  & 562 & $1.2 \times 10^{-6}$ & 0.28 & 44.58 (44.53)\\
Low $L_{\rm X}$ RASS-AGN sample & $0.16 <z<0.36$ &$43.6\la \log L_{\rm X} \le 44.3$& 990  & $5.8 \times 10^{-5}$ & 0.24 & 43.95 (44.16) \\
\enddata
\tablenotetext{a}{X-ray luminosity is measured in the rest-frame energy range 0.1-2.4 keV.}
\tablenotetext{b}{The symbol ``$\la$'' is used in case of a ``soft'' luminosity boundary and 
  indicates the 10th (for the lower bound) or the 90th (for the upper bound) percentile from the lowest luminosity object 
  in the sample. The symbols $\le$ and $<$ are used to indicate a hard luminosity boundary we impose.}
\tablenotetext{c}{The the logarithm of the mean $L_{\rm X}$ is followed by parenthesized median values.}
\end{deluxetable*}

 In calculating the AGN-LRG CCFs and the LRG ACF, we use the classic 
\citet{davis_peebles83} estimator, in a two-dimensional (2D) grid in the
$(r_{\rm p},\pi)$ space, where $r_{\rm p}$ is the projected distance and
$\pi$ is the line-of-sight separation (both in comoving coordinates): 
\begin{equation}
\xi(r_{\rm p},\pi)=\frac{D_1D_2(r_{\rm p},\pi)}{D_1R_2(r_{\rm p},\pi)}-1,
\label{eq:dp83}
\end{equation}
in which the subscripts 1 and 2 identify the sample, $D_1D_2$ is the number
of pairs between real samples, and $D_1R_2(r_{\rm p},\pi)$ is the number
of pairs between real sample 1 and random sample 2.  
For the AGN-LRG CCF, the sample 1 is the AGN sample and
the sample 2 is the LRG sample, while for the LRG ACF, both 
sample 1 and sample 2 are the same LRG sample.  We use Eq. \ref{eq:dp83}
rather than that proposed by \citet{landy_szalay_1993} because Eq. \ref{eq:dp83}
requires a random sample for only sample 2 (in our case, LRGs). This is
important due of the difficulty in generating a random sample for
a RASS-based AGN catalog. {\sl ROSAT} is sensitive to soft X-rays, 
which are subject to absorption by neutral hydrogen. The variation 
in the column density across the sky due to neutral gas in our galaxy, 
combined with the diversity of the X-ray spectra of AGN, makes it very difficult
to model the spatial variation in the detection limit of the AGNs,
which is essential in generating a random sample.

 We then calculate the projected correlation function:
\begin{equation}
w_{\rm p}(r_{\rm p})=2 \int_0^{\pi_{\rm max}}\xi(r_{\rm p},\pi)d\pi,
\end{equation}
where the upper bound of the integral ($\pi_{\rm max}$) is determined by 
the saturation of the integral. \chg{We take $\pi_{\rm max}=80$ and
$40h^{-1}$ Mpc for the LRG ACF and AGN-LRG CCF respectively. \alc{The reasoning 
for these choices is explained in paper I.}}

In paper I, we further calculate the {\em inferred} AGN ACF, $w_p(AGN|AGN)$, 
from the AGN-LRG CCF $w_p(AGN|LRG)$ and LRG ACF $w_p(LRG|LRG)$, assuming a 
linear biasing scheme:
\begin{equation}
 w_p(AGN|AGN) = \frac{\left[w_p(AGN|LRG)\right]^2}{w_p(LRG|LRG)}.
\label{eq:acf_infer}
\end{equation}
 We then fit $w_p(AGN|AGN)$ with a power-law form 
and derive the AGN bias parameters and typical masses of the DMH in 
which the AGNs reside, based on the power-law fits. 

 In this paper, {instead of using the inferred AGN ACF as in Paper I, 
we directly model the CCF, $w_p(AGN|LRG)$, with an HOD analysis, as $w_p(AGN|LRG)$ 
has been directly derived from the observations and does not depend on a linear 
biasing approximation.}  The comparison of results from paper I and this paper are 
discussed below in Sect. \ref{sec:comp_paperI}.  

\section{Modeling the Halo Occupation Distribution}
\label{sec:halomodel}
\subsection{Model Ingredients}
\label{sec:model_ing}

  In performing the HOD modeling, we consider that galaxies and AGNs are 
associated with DMHs, the mass function of which (per comoving volume per $dM_{\rm h}$) 
is denoted by $\phi(M_{\rm h})\,dM_{\rm h}$, where $M_{\rm h}$ is the dark matter halo mass. 
We use $\phi(M_{\rm h})$ based on \citet{st99}, which is in good agreement
with later analyses by \citet{sheth01} and \citet{jenkins01}. 
A DMH may contain one or more galaxies and/or AGNs that are 
included in our samples. 
We then model the two-point projected correlation function 
as the sum of two terms,
\begin{equation}
  w_{\rm p}(r_{\rm p})=w_{\rm p,1h}(r_{\rm p}) + w_{\rm p,2h}(r_{\rm p}),
\label{eq:wrp_1h2h}
\end{equation}
where the terms are:
\begin{itemize}
\item the 1-halo term (denoted by a subscript ``1h''), where both of the objects 
  occupy the same DMH, and
\item the 2-halo term (denoted by a subscript ``2h''), where each of the pair of 
   objects occupies a different DMH.   
\end{itemize}

Recent articles prefer to define $\xi_{\rm 1h}$ such that
the total 3D correlation function is expressed by 
$\xi=(1+\xi_{\rm 1h})+\xi_{\rm 2h}$ \citep{tinker05,zheng05,blake08}, 
instead of $\xi=\xi_{\rm 1h}+\xi_{\rm 2h}$, as used in older articles. 
This is because $1+\xi$ 
represents a quantity that is proportional to the number of pairs, and allows one 
to express each of the 1- and 2-halo terms consistently. In this 
case, the number of pairs is $\propto 1+\xi=[1+\xi_{\rm 1h}]+[1+\xi_{\rm 2h}]$.  
In this new convention, our $w_{\rm p,1h}$ represents the projection of $1+\xi_{\rm 1h}$ 
rather than $\xi_{\rm 1h}$}, i.e.,
\begin{equation} 
w_{\rm p,1h}(r_{\rm p}) = \int_{-\infty}^\infty \left[1+\xi_{\rm 1h}\left(r_{\rm p},\pi\right)\right] d\pi.
\end{equation}

 Similarly, we express the power spectrum of the distribution of the objects
in terms of the 1- and 2-halo term contributions:
\begin{equation}
P(k) = P_{\rm 1h}(k)+P_{\rm 2h}(k),
\end{equation}
with 
\begin{eqnarray}
w_{\rm p,1h}(r_{\rm p})&=&\int \frac{k}{2\pi}P_{\rm 1h}(k) J_0(kr_{\rm p})dk \nonumber \\ 
w_{\rm p,2h}(r_{\rm p})&=&\int \frac{k}{2\pi}P_{\rm 2h}(k) J_0(kr_{\rm p})dk, 
\label{eq:wp_pk}
\end{eqnarray}
where $J_0(x)$ is the zeroth-order Bessel function of the first kind.  

 In the case of modeling the ACF of a sample,
the 2-halo term of the power spectrum can be approximated by 
\citep{cooray_sheth02}
\begin{equation}
P_{\rm 2h}(k) \approx b^2 P_{\rm lin}(k,z),
\label{eq:p2h_acf}
\end{equation} 
where $b$ is the bias parameter of the sample, which we model as 
\begin{equation}
b = \frac{\int b_{\rm h}(M_{\rm h})\langle N\rangle (M_{\rm h}) \phi(M_{\rm h})dM_{\rm h}}
         {\int \langle N\rangle (M_{\rm h}) \phi(M_{\rm h})dM_{\rm h}}.
\end{equation} 
For the linear power spectrum, $P_{\rm lin}(k,z)$, we use the primordial power spectrum with $n_{\rm s}=1$  
and a transfer function calculated using the fitting formula of \citet{eisenstein_hu98} under our assumed 
cosmology (Sect. \ref{sec:intro}). For the mass-dependent bias parameter of DMHs, $b_{\rm h}(M_{\rm h})$, we 
use Eq. (8) of \citet{sheth01} with recalibrated parameter values by \citet{tinker05} (see their Appendix A).

 In calculating the 1-halo term, the usual assumption is that the radial 
distribution of the involved objects that 
are not at the center of the halo -- objects that are satellites -- follows the 
mass profile of the DMH itself, i.e., there is no internal 
biasing. Following \citet{zheng09}, whose results will be used later in this work, 
we use the \citet{nfw97} (NFW) profile for the DMH density distribution, \chg{which 
is still a popular choice for representing DMH profiles, although other parametrizations exist 
in the literature \citep[e.g.][]{knollmann08,stadel09}.}  
We express the Fourier transform of the 
NFW profile of the DMH with mass $M_{\rm h}$, normalized such that volume integral up to the virial
radius is unity,  by $y(k,M_{\rm h})$.  Again, in the case of the ACF,
\begin{eqnarray}
P_{\rm 1h}(k)&=&\frac{1}{(2\pi)^3n^2}\int \phi(M_{\rm h})[\langle N_{\rm c}N_{\rm s}\rangle(M_{\rm h})
                y(k,M_{\rm h})+\nonumber\\
             & &\langle N_{\rm s}(N_{\rm s}-1)\rangle(M_{\rm h})|y(k,M_{\rm h})|^2]dM_{\rm h},
\label{eq:p1h_acf}
\end{eqnarray}
where  $N_{\rm c}$ and $N_{\rm s}$ are the numbers of the objects in the sample per DMH 
as a function of $M_{\rm h}$ for those that are at the center of the halos (central) 
and those that are off center (satellites) respectively, while 
$n=\int \langle N\rangle(M_{\rm h})\phi(M_{\rm h})dM_{\rm h}$ is their overall space density 
\citep{hamana04,seljak00}.  

 The HOD model is calculated using a set of software we have developed, 
based partially on a code developed by J. Peacock for the use in 
\citet{peacock00} and \citet{phleps06}. Our software includes a number of improvements over their 
code, most notably by including the application of the  HOD modeling to a CCF.
We note, however, that in our analysis we do not include some recent 
improvements in modeling the 2-halo term  \citep{zheng05,tinker05,zheng09}. 
The neglected factors include the convolution of the DMH profile to the 2-halo term 
and the scale-dependent bias. In addition, if a pair of objects are closer than the sum of the virial radii 
of their parent halos, the pair should not be counted in the 2-halo term; this 
effect is neglected. These improvements have enabled those authors 
to make accurate modelings of, e.g., the ACF of SDSS galaxies, where the correlation 
functions can be measured to a few percent level at 
$r_{\rm p}\approx 1 h^{-1}$ Mpc. However, as AGNs have a much 
smaller number density, the errors on our CCF measurements are 
$\gtrsim 20$\% (paper I) at this $r_{\rm p}$, where these improvements are important. 
Therefore the effects of these recent improvements by other authors 
are estimated to be within the 1$\sigma$ measurement  errors of the CCF.
         
\subsection{The HOD modeling of the AGN-LRG CCF}

  In this subsection, we explain the application of the  HOD modeling to the 
CCF between AGNs and LRGs (or more generally, between any two different populations).  
Using the HOD of the LRGs as a template,
which has been accurately constrained by Z09 using the LRG ACF measured by
\citet{zehavi05b}, we constrain the HOD of the AGNs by fitting to 
the AGN-LRG CCF. A similar approach has been made by Z09 to model the CCF 
between LRGs and $L_*$ galaxies, where they assumed that the LRGs are at the 
halo centers and the $L_*$ galaxies are satellites {in the 1-halo
term calculation}.
{Since LRGs occupy the centers of almost all massive DMHs  
($\log M_{\rm h}\ga 14\,[h^{-1}M_{\sun}]$, Z09), 
this is a good approximation for their LRG-$L_*$ galaxy CCF as well as for our AGN-LRG CCF.
However, for completeness,  we develop a formulation of the  HOD modeling
that includes more general cases.} 
Hereafter, the quantities for AGNs are represented by a subscript ``A'', 
LRGs by ``G'' (representing {\it galaxies}), and CCF between the two by ``AG''.  
We denote the LRG HODs at the halo center and of satellites by 
$\langle N_{\rm G,c}\rangle (M_{\rm h})$ and $\langle N_{\rm G,s}\rangle (M_{\rm h})$, 
respectively. Now 
$\langle N_{\rm G}\rangle (M_{\rm h})=\langle N_{\rm G,c}\rangle (M_{\rm h})+\langle N_{\rm G,s}\rangle (M_{\rm h})$. 
Likewise, the HODs of the AGNs at the halo centers, of satellites, and the sum of 
the two are denoted by 
$\langle N_{\rm A,c}\rangle (M_{\rm h})$, $\langle N_{\rm A,s}\rangle (M_{\rm h})$ and 
$\langle N_{\rm A}\rangle (M_{\rm h})$. 

 In our approximate treatment, the two halo term can be expressed as follows. 
Let the linear bias parameters of the LRGs and AGNs be $b_{\rm G}$ and $b_{\rm A}$ 
respectively:
\begin{eqnarray}
b_{\rm G}&=&\frac{\int b_{\rm h}(M_{\rm h})\langle N_{\rm G}\rangle (M_{\rm h})\phi(M_{\rm h})dM_{\rm h}}
 {\int \langle N_{\rm G}\rangle(M_{\rm h})\phi(M_{\rm h})dM_{\rm h}}\\
b_{\rm A}&=&\frac{\int b_{\rm h}(M_{\rm h})\langle N_{\rm A}\rangle (M_{\rm h})\phi(M_{\rm h})dM_{\rm h}}
        {\int \langle N_{\rm A}\rangle (M_{\rm h})\phi(M_{\rm h})dM_{\rm h}}.
\label{eq:bias_A}
\end{eqnarray}
Then the two halo term of the power spectrum corresponding to the CCF 
(cross power spectrum) can be expressed by
\begin{equation} 
P_{\rm AG, 2h}(k) \approx b_{\rm A}b_{\rm G}  P_{\rm lin}(k). 
\label{eq:p_ccf_2h}
\end{equation}
 
 The 1-halo term is composed of three terms:
\begin{eqnarray}
P_{\rm AG,1h}(k)&=&\frac{1}{(2\pi)^3n_{\rm A}n_{\rm G}}\int\phi(M_{\rm h})\times\nonumber\\
        & &[\langle N_{\rm A,c}N_{\rm G,s}+N_{\rm A,s}N_{\rm G,c}\rangle(M_{\rm h})\,\,y(k,M_{\rm h})+\nonumber\\
        & &\langle N_{\rm A,s}N_{\rm G,s}\rangle(M_{\rm h})\,\,|y(k,M_{\rm h})|^2]\, dM_{\rm h},
\label{eq:pag1h}
\end{eqnarray}  
\chg{provided that no object is in common between the AGN and LRG samples (which 
is the case in our CCF). In this case, the relation}
\begin{equation}
\langle N_{\rm A,x}N_{\rm G,y}\rangle(M_{\rm h})=
\langle N_{\rm A,x}\rangle(M_{\rm h})\langle N_{\rm G,y}\rangle(M_{\rm h})
\label{eq:nxngmean}
\end{equation}    
holds, where the subscripts x and y represent any combination of the subscripts s and c\chg{,
except for the case both are c's}. This relation is exact even if \chg{$\langle N \rangle\lesssim 1$, 
where one has to use the exact Poisson distribution for the probability distribution of having exactly 
$N$ objects in a halo (sometimes called a sub-Poissonian case).}
The relation $\langle N_{\rm A,c}N_{\rm G,s}+N_{\rm A,s}N_{\rm G,c}\rangle=
\langle N_{\rm A,c}N_{\rm G,s}\rangle+\langle N_{\rm A,s}N_{\rm G,c}\rangle$ also holds.   

\chg{Even in \alc{cases where there are objects in common between the two samples, the above discussion 
  does not change for the central-satellite pairs, because the two objects are surely different.} 
  However, the existence of the common objects makes assumptions behind the third term in the 
  integrand of Eq. \ref{eq:pag1h} invalid. A detailed discussion of this case is beyond the scope 
  of the present paper, and will be made in a future paper.}

\subsubsection{The LRG HOD}
\label{sec:lrghod}

 The HOD of the LRGs has been extensively studied by Z09 based on the ACF measurements
of the LRGs by \citet{zehavi05b} using SDSS Data Release (DR) 3. 
Their $\langle N_{\rm G,s}\rangle(M_{\rm h})$ model
has essentially five parameters, which are the values at five $M_{\rm h}$ points, and 
they have spline-interpolation values between these points. 
They take an elaborate parametrization   
of $\langle N_{\rm G,c}\rangle(M_{\rm h})$, which is an integration of  a 
luminosity-dependent smoothed step-function, where the value increases from 0 
(at low $M_{\rm h}$) to 1 (at high $M_{\rm h}$), 
with a transition following a luminosity-dependent error function. 
  
 In paper I, we re-calculate the LRG ACF for the $0.16<z<0.36$, $-23.2<M_g<-21.2$ sample 
based on DR4+, using $\approx$46,000 LRGs, instead of $\approx$30,000 LRGs from DR3 used 
by \citet{zehavi05b}. We obtain practically the identical ACF to that calculated by 
\citet{zehavi05b}. 
\chg{Since we want to use exactly the same model and the same LRG
  sample between the LRG ACF and the LRG-AGN CCF,\alc{ we derive the best-fit 
LRG
  HOD to our DR4+ sample using our software. However,
  instead of newly exploring the full parameter space, we take
  advantage of the analysis in Z09 to find the HODs that give the best fit
  to our data by tweaking their LRG HODs as follows.}}
Fig. 1 (b) of Z09 shows the $\Delta \chi^2=4$ upper and lower bounds of the HODs 
separately for the central and satellite LRGs.   
For the satellites, we start with their $\Delta \chi^2=4$ 
upper and lower bounds, denoted by $N_{\rm Z,s}^{\rm u}$ and $N_{\rm Z,s}^{\rm l}$ 
respectively.  We then interpolate between these curves:
\begin{equation}
\langle N_{\rm G,s}\rangle(M_{\rm h}) = (1-f)\langle N_{\rm Z,s}^{\rm l}\rangle (M_{\rm h})+
f\langle N_{\rm Z,s}^{\rm u}\rangle (M_{\rm h}), \nonumber\\
\end{equation} 
where $f$ is a weight parameter in the linear interpolating procedure.

For the central LRGs, we shift their central HOD ($N_{\rm Z,c}$) horizontally 
by $d$ in $\log M_{\rm h}$:
\begin{equation}
\langle N_{\rm G,c}\rangle(M_{\rm h}) = \langle N_{\rm Z,c}\rangle(M_{\rm h}10^d)\nonumber\\
\end{equation}
We tweak $N_{\rm G,c}$ horizontally because of the 
$\langle N_{\rm G,c}\rangle(M_{\rm h}) \leq 1$ constraint.  In calculating 
$\langle N_{\rm s}(N_{\rm s}-1)\rangle(M_{\rm h})$ of the LRG ACF based on the above HODs, 
we calculate the mean value using the exact Poisson statistics in the small
number case ($\langle N_{\rm s}\rangle\le 10.0$), while we use 
$\langle N_{\rm s}\rangle(\langle N_{\rm s}\rangle-1)$ 
for larger values. As a whole, our model has two ``tweak'' parameters, $f$ and $d$.

 We calculate a series of model ACFs at $z=0.28$ (the average redshift of our sample, 
see Table \ref{tab:samp}) in a parameter space grid \chg{exhaustively over a rectangular 
area in the $(f,d)$ space that is large enough to include all the region with 
$\Delta \chi^2\lesssim 5$. The grid spacings are 0.03 and 0.003 for 
$f$ and $d$ respectively.}  

 Then we fit the model to the data by minimizing $\chi^2$, taking the 
correlation of errors and the total number density of LRGs ($n_{\rm LRG}$) into account:
\begin{eqnarray}
\chi^2 &=& \sum_{ij}\{[w_{\rm p}(r_{{\rm p},i})-w_{\rm p}^{\rm mdl}(r_{{\rm p},i})]M^{-1}_{ij}\times\nonumber\\
       & & [w_{\rm p}(r_{{\rm p},j})-w_{\rm p}^{\rm mdl}(r_{{\rm p},j})]\} +\nonumber\\
       & &  (n_{\rm LRG}-n_{\rm LRG}^{\rm mdl})^2/\sigma_{n_{\rm LRG}}^2,
\label{eq:acf_chi2}
\end{eqnarray}
where the quantities from the model are indicated by a superscript ``mdl'', $M$ is the covariance 
matrix, and $\sigma_{n_{\rm LRG}}$ is the 1$\sigma$ error of the LRG number density. 
\chg{The covariance matrix and $\sigma_{n_{\rm LRG}}$ are estimated using the jackknife 
resampling method (see paper I). During the minimization process, the models 
$w_{\rm p}^{\rm mdl}(r_{{\rm p},i})$ are calculated by 
interpolating from the four nearest grid points in the parameter space computed above. 
The minimization is performed using the MINUIT 
package\footnote{\url{http://wwwasdoc.web.cern.ch/wwwasdoc/minuit/minmain.html}} 
distributed as a part of the CERN program library.} Including the number density term in 
$\chi^2$ gives important parameter constraints, not only on the overall normalization, 
but also the shape. This is because $N_{\rm G,c}$, by definition, has the absolute maximum 
value of unity (as there can not be more than one central galaxy in a halo), and for the 
LRGs, the value saturates at 1 for $M_{\rm h}\gtrsim 10^{\rm 14}h^{-1}\,M_{\odot}$. 
With this constraint, $n_{\rm LRG}$ becomes sensitive to the combination of the relative normalizations 
between $N_{\rm G,c}$ and $N_{\rm G,s}$ and the  ``cut-off'' mass of $N_{\rm G,c}$.  
For the LRG sample defined in paper I, $n_{\rm LRG}=(9.56\pm 0.13)\times 10^{-5}h^{-3}$ Mpc$^{-3}$.
In performing the fits, we use only the data points that are dominated by 
either the 1-halo or 2-halo terms, excluding the transition region
 ($0.46<r_{\rm p}[h^{-1}{\rm Mpc}]<2.8$). 
In this transition region, our model ACF becomes inaccurate due to our approximations, 
especially in neglecting the halo-exclusion effect in the 2-halo term, as described 
above in Sect 3.1. We use the data points down to $r_{\rm p}=0.2\,[h^{-1}{\rm Mpc}]$, 
which is slightly below the lower $r_{\rm p}$ limit used in paper I.  Our measurements 
to this scale are consistent with those measure by \citet{mesjedi06}. 
 
\begin{figure}
\begin{center}
\includegraphics[width=\hsize]{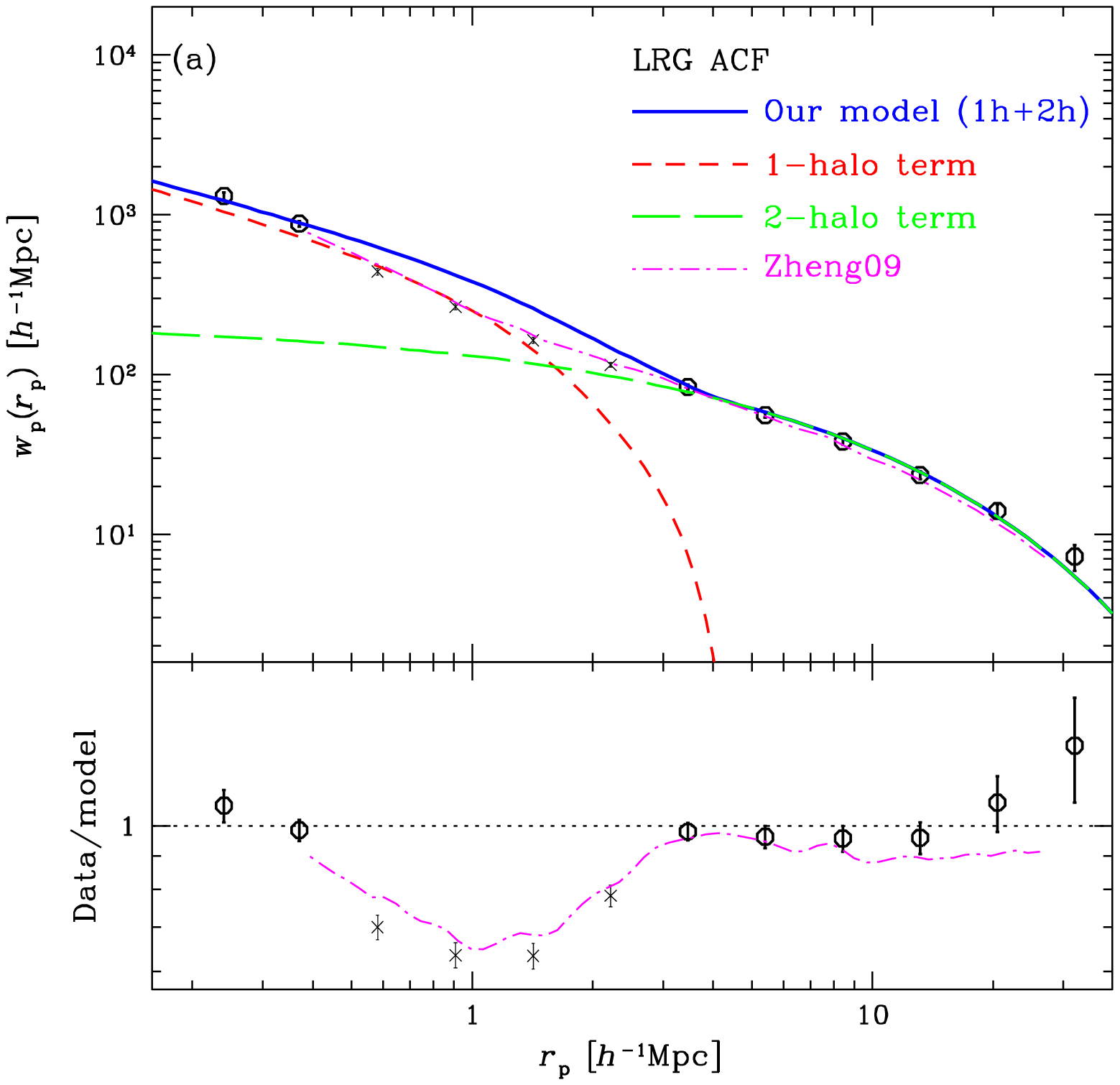}
\resizebox{\hsize}{!}{
  \centering
  \includegraphics[height=0.51\hsize]{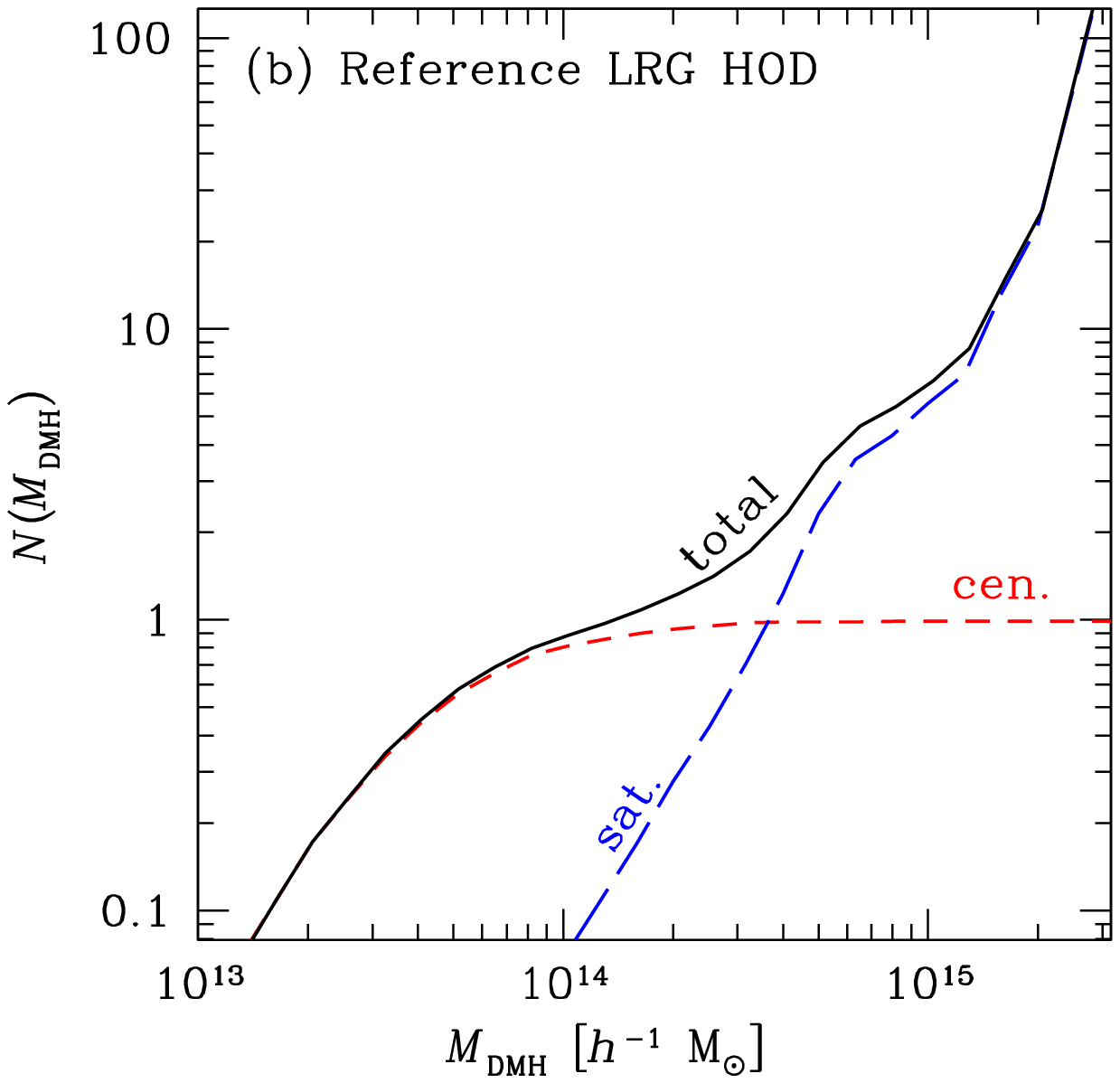}
  \includegraphics[height=0.50\hsize]{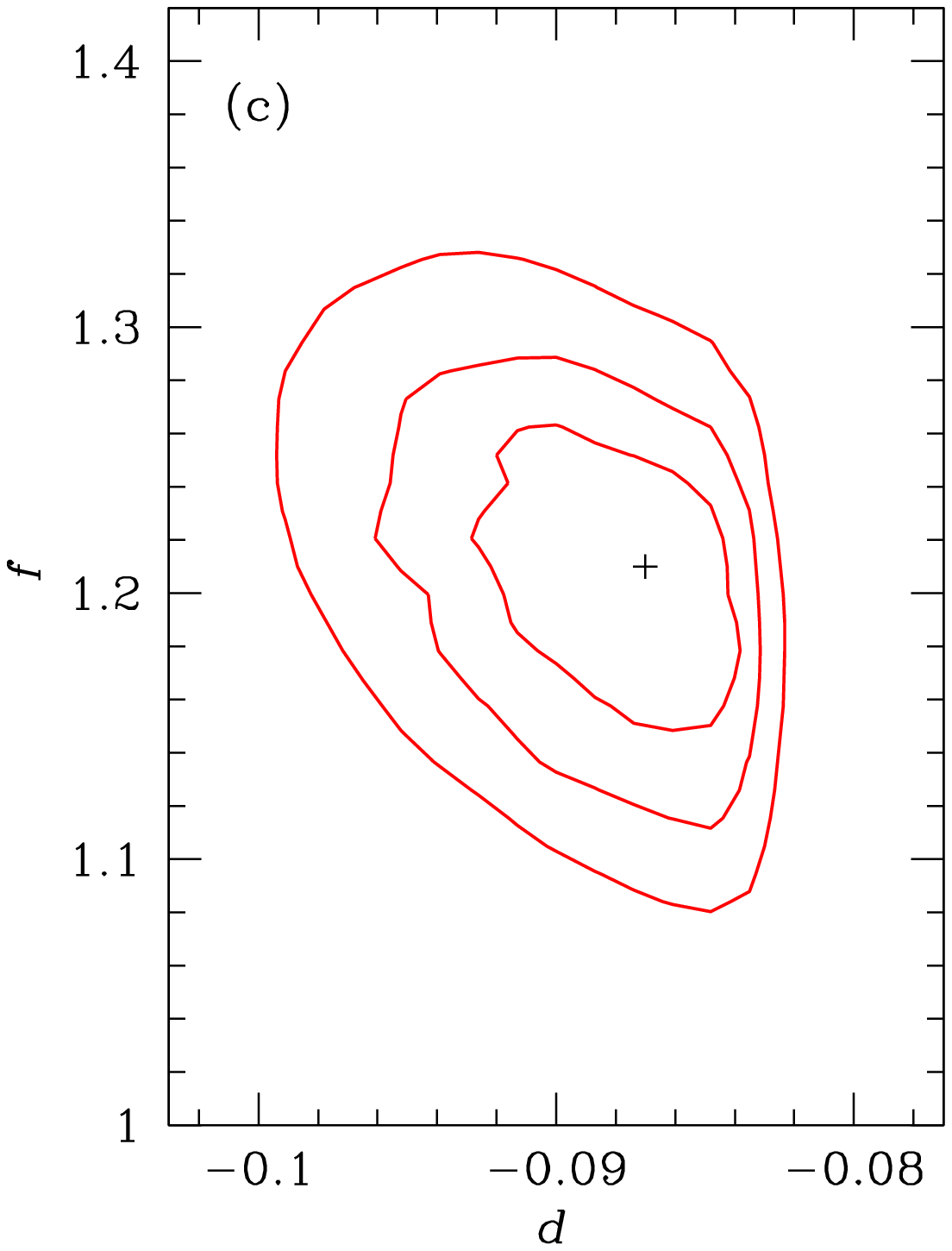}
}
\end{center}
 \caption{(a) The ACF of the luminous red galaxy (LRG) sample and the best-fit HOD 
    model (upper panel) with residuals (lower panel). The data points and error bars show 
    our measurements of the LRG ACF. 
    The circles show the data points used for the fit, while the crosses show those excluded. The 
    dashed (green), long-dashed (green) and thick-dashed lines show the 1-halo
    term, 2-halo term, and the total respectively. The dot-dashed (magenta) curve represents  
    the best-fit model by Z09. Our simplified model fits well in the regimes where the
    1-halo term and 2-halo terms dominate. 
    (b) Our best fit HODs for the central LRGs (red dashed line), satellite LRGs 
    (blue long-dashed line),  and the total. (c) Confidence contours in the $d$-$f$ space 
    (see text for details). The confidence levels are $\Delta \chi^2=1.0, 2.3$ and 4.6. The cross 
    shows the best-fit values.
   }
   \label{fig:lrgacfhod}
\end{figure}

 Our best fit model has `tweak' parameter values of $f=1.21 (1.12; 1.29)$ and 
$d = -0.087 (-0.095; -0.083)$, where the 68\% confidence ranges for two interesting 
parameters ($\Delta \chi^2<2.3$) are given in parentheses. Our HODs are 
slightly higher than those measured by Z09, probably because 
we include a data point at $r_{\rm p}=0.2\,[h^{-1}{\rm Mpc}]$ and we use 
a slightly different linear power spectrum.  
The best-fit model $w_{\rm p}(r_{\rm p})$ is compared with the 
observation as well as the Z09 model in Fig.~\ref{fig:lrgacfhod}(a). The corresponding 
HODs ($\langle N_{\rm G,s}(M_{\rm h})\rangle$, $\langle N_{\rm G,c}(M_{\rm h})\rangle$ and 
the total) are shown in Fig.~\ref{fig:lrgacfhod}(b), and confidence contours in the
($d$,$f$) space are shown in Fig.~\ref{fig:lrgacfhod}(c). 

\placefigure{fig:lrgacfhod}

\subsubsection{The AGN HOD}
\label{sec:agnhod_ex}

With the LRG HOD in hand, we come to our main purpose of obtaining constraints on the 
HOD of the AGNs in the RASS-SDSS survey. 

\chg{As a parametrized form of the AGN HOD, we first try a simple truncated power-law
form, assuming that all the AGNs are satellites (Model A):

\begin{equation}
\langle N_{\rm A,s}\rangle \propto M_{\rm h}^\alpha\Theta (M_{\rm h}-M_{\rm cr}),\,\,\langle N_{\rm A,c}\rangle=0   
\label{eq:agnhod}
\end{equation}   
where $\Theta(x)$ is the step function ($=1$ at $x\geq 0$; $=0$ at $x<0$), \chg{$M_{\rm cr}$ is a critical
DMH mass below which the HOD is zero, and $\alpha$ is a power-law slope of the HOD above $M_{\rm cr}$.}
\alc{While formally we are assuming that all AGNs are satellites, the HOD constraints obtained
from this assumption can be applied fairly to the sum of the central and satellite AGNs
in cases where the AGN activity in the central galaxy of high mass 
$\log\,M_{\rm h} [h^{-1}M_\sun]\gtrsim 13.5$ DMHs is suppressed. 
For simplicity of explanation, we visualize 
the HOD of the central LRG as a step function with a step at $\log\,M_{\rm h} [h^{-1}M_\sun]=13.5$, 
and refer the DMH mass roughly above (below) this cut as higher (lower) mass. 
There is no contribution to the 1-halo term at lower mass DMHs, since the 1-halo term requires 
both AGNs and LRGs to reside at the same DMH mass. Since there 
is in practice no distinction between central and satellite AGNs in the 2-halo term 
(Eq. \ref{eq:p_ccf_2h}), whether the AGNs in low mass DMHs are satellites or central makes 
no difference to the CCF, and indeed this model can fairly accommodate a case where a low-mass 
halo represents one galaxy (halo) at the center.  At higher masses,
among the three terms in Eq. \ref{eq:pag1h} the satellite AGN-central LRG pairs dominate 
the 1-halo term, because, as we show later, the present HOD analysis is sensitive at 
$\log\,M_{\rm h} [h^{-1}M_\sun]<14.0$, where there are in practice no 
satellite LRGs (Fig. \ref{fig:lrgacfhod}), and  therefore the central AGN-satellite LRG
term as well as the satellite AGN-satellite LRG term are negligible. 
Thus Eq. \ref{eq:agnhod} represents the sum of the central and satellite AGN HOD if there is no 
central AGN at higher masses.

 
The underlying assumption in Model A is partially motivated by 
observations of AGN ``downsizing'' \citep{ueda03,hasinger08,ebrero09a,yencho09}, where the number 
density of high luminosity AGNs peaks earlier in the history of the universe and
drops rapidly towards low redshift, while lower luminosity AGN activity
peaks later. (We note, however,  that a recent work by \citet{aird10} reports 
that this trend might be weaker than those reported previously.)  
One possible implication of this trend is that black holes
at the centers of massive galaxies occupying the centers of (higher mass) DMHs 
stopped accreting long before $z\sim 0.3$, while accretion occurs 
more frequently in lower mass galaxies in satellites or at the centers of lower 
mass halos in the redshift range of our sample. Thus model A can be a demonstrative case of 
this scenario. Additionally, central galaxies of massive halos are mostly early-type luminous 
galaxies, where AGN activity is reported to be suppressed \citep{schawinski10}. 

In Sect. \ref{sec:modelB} we consider in detail other models (Models B and C) which uses well-studied 
galaxy HODs (central plus satellites) as templates and treat the separate central and satellite HODs 
explicitly.}}

 Using a fixed set of $\langle N_{\rm G,c}\rangle (M_{\rm h})$ and 
$\langle N_{\rm G,s}\rangle (M_{\rm h})$  
derived in the previous subsection and the formulations developed earlier in this section, 
we calculate the expected $w_{\rm p,AG}(r_{\rm p})$ in the two parameter ($M_{\rm cr}$,$\alpha$) model of 
$\langle N_{\rm A}\rangle (M_{\rm h})$. 
\chg{Due to much smaller errors of the LRG ACF compared with those in
  the AGN-LRG CCF, we fixed the tweak parameter $f$ and $d$ at the
  best-fit values during the fits to the CCF. Shifting the tweak
  parameters to any point on the $\Delta \chi^2=2.3$ contour in
  Fig.~\ref{fig:lrgacfhod}(c) causes a shift of only $\Delta
  \chi^2<0.05$ to the same AGN HOD model.  This justifies the use of
  fixed LRG HODs during the fit to the CCF.}

 We calculate a series of model CCFs in a parameter grid at $z=0.28$ \chg{with the spacings of 0.1 and 0.05 
for $\log\,M_{\rm cr}$ and $\alpha$, respectively, over the 
rectangular region that we are 
interested in (see confidence contours in the next section).}   
While the mean redshift of the AGN samples range from $z=0.24$ to $z=0.28$, the 1-halo 
and 2-halo terms vary by only $\approx 2$\% and $\approx 0.1$\% respectively between
these redshifts, justifying our single redshift calculations.    
 We search for the  best-fit model by minimizing the correlated $\chi^2$:
\begin{eqnarray}
\chi^2 &=& \sum_{ij}\{[w_{\rm p,AG}(r_{{\rm p},i})-w_{\rm p,AG}^{\rm mdl}(r_{{\rm p},i)}]M^{-1}_{ij}\times\nonumber\\
       & & [w_{\rm p,AG}(r_{{\rm p},j})-w_{\rm p,AG}^{\rm mdl}(r_{{\rm p},j})]\},
\label{eq:chi2ccf}
\end{eqnarray}
\chg{where the errors and covariance matrix are estimated using the jackknife resampling
method (paper I).}
 Unlike in the case of LRGs, here we do not include the number density term in 
determining $\chi^2$, because \chg{the CCF depends only on $\alpha$ and $M_{\rm cr}$
in Eq. \ref{eq:agnhod} but not the normalization. Thus the CCF constraints on  
$\alpha$ and $M_{\rm cr}$ and the density constraints on the normalization can be 
separated.} After the constraints of these two parameters are made, the normalization can be 
determined by matching to the number density of the AGNs.  We use only data points with 
$r_{\rm p}>0.3 h^{-1}$Mpc, as smaller scale bins contain only a small number of AGN-LRG pairs 
($<15$ pairs for the total and much less in the high and low $L_{\rm X}$ RASS-AGN samples), 
which would not warrant the applicability of the $\chi^2$ statistic. Unlike in the case of the LRG ACF, 
we use all the data points in $0.3\la r_{\rm p}[h^{-1}{\rm Mpc}]\la 40$. Due to the 
smaller mass for the DMH occupied by the AGN compared to satellite LRGs, the transition 
between 1-halo and 2-halo dominated regimes is very narrow for the AGN-LRG CCFs, as 
will be seen in the next section. Even within this small transition region, the effects of the major source 
of inaccuracy in our approximation, i.e., halo-exclusion in the 2-halo term,
is estimated be a few times smaller than our CCF measurement errors, {based on 
the comparison of our LRG 2-halo term and that calculated by Z09}. 
 
\section{Results from the  HOD Modeling}
\label{sec:results}

\subsection{Fit Results}
\label{sec:results_fit}

\begin{figure*}
  \centering
 \includegraphics[height=0.46\hsize]{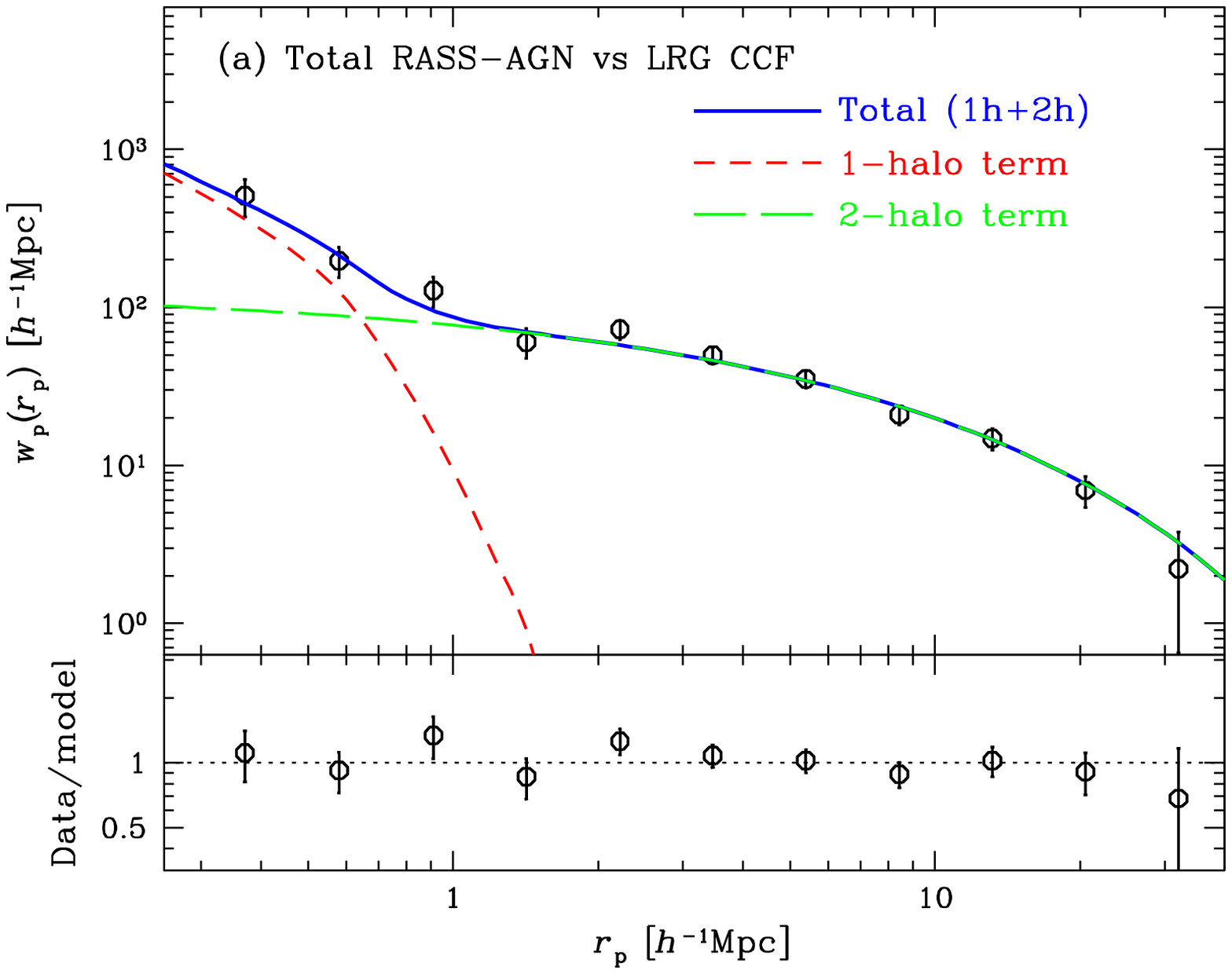}
 \includegraphics[height=0.46\hsize]{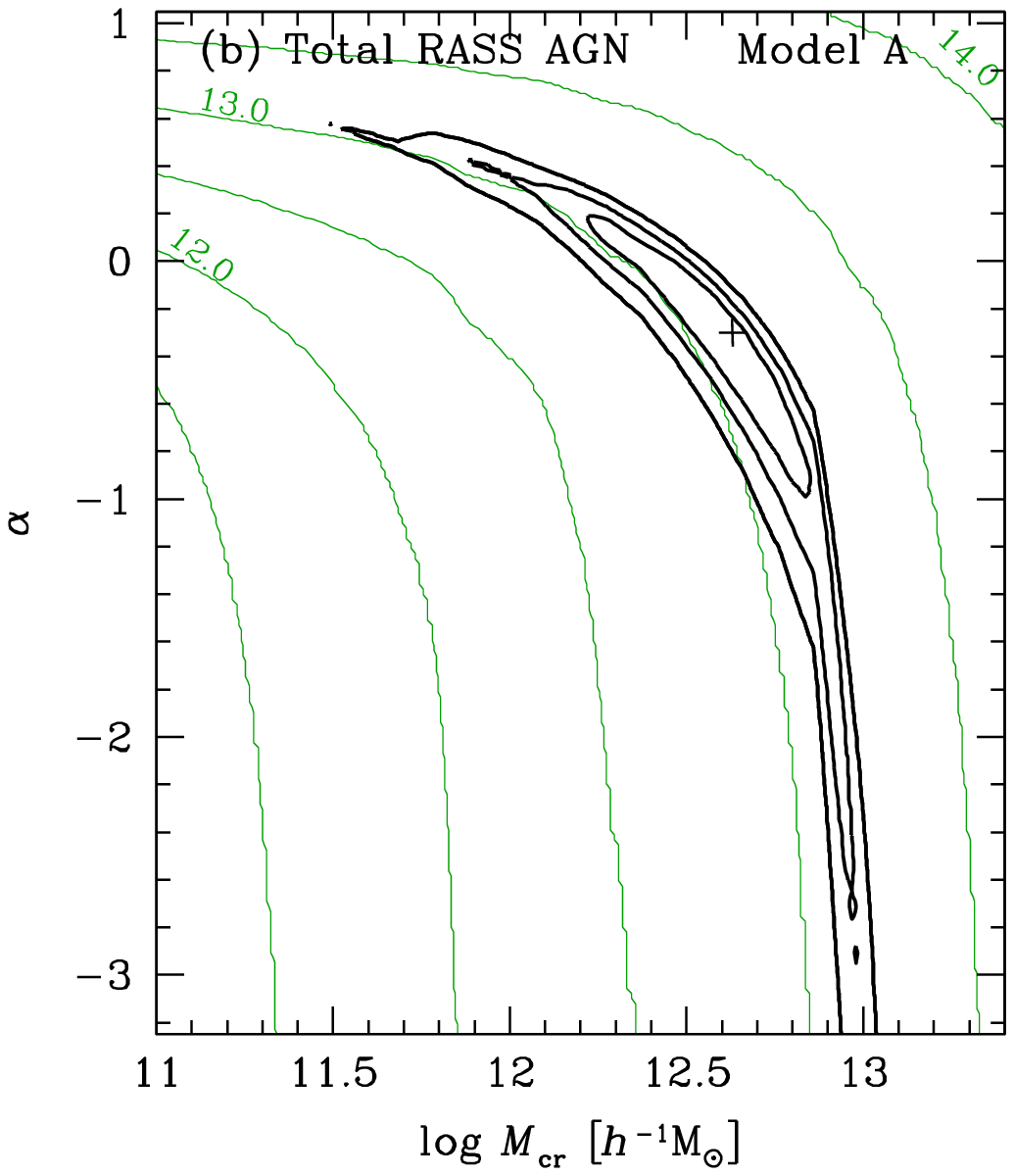}
 \caption{(a) The AGN-LRG CCF (for the total RASS-AGN sample) and our {best-fit} 
   HOD model for Model A. The symbols and line styles are the same as in Fig. \ref{fig:lrgacfhod}(a).
    (b) The confidence contours of the Model A parameters ($\log\,\,M_{\rm cr}$,$\alpha$).
    The confidence contours (thick contours) are at the $\Delta \chi^2=1.0,2.3$ and $4.6$ levels. 
    The thin-contours (green) show the mean halo mass $\log\, \langle  M_{\rm h}[h^{-1}{\rm M_\sun}]\rangle$ 
    derived from the model parameters. Every other contour level is labeled. The contours show that 
    the parameter constraints are almost along a constant $\log\, \langle M_{\rm h}\rangle$ line.}
   \label{fig:tot_ccf_con}
\end{figure*}


 Using the methods described above, we obtain constraints on the AGN HOD parameters
$\log\,M_{\rm cr}$ and $\alpha$ in Eq. \ref{eq:agnhod}. For each point in this two parameter space, we can 
also calculate the mean dark matter halo mass occupied by the AGN sample,
\begin{equation}
\langle M_{\rm h}\rangle=\frac{\int M_{\rm h} \langle N_{\rm A}\rangle (M_{\rm h}) \phi(M_{\rm h})dM_{\rm h}}
                      {\int \langle N_{\rm A}\rangle (M_{\rm h}) \phi(M_{\rm h})dM_{\rm h}},
\label{eq:meanmass}
\end{equation}
as well as the effective bias parameter of AGNs $b_{\rm A}$ (Eq. \ref{eq:bias_A}).

 Figure \ref{fig:tot_ccf_con}(a) compares our AGN-LRG CCF for the total RASS-AGN sample
and our best-fit model. 

\begin{figure}
 \centering
 \includegraphics[width=\hsize]{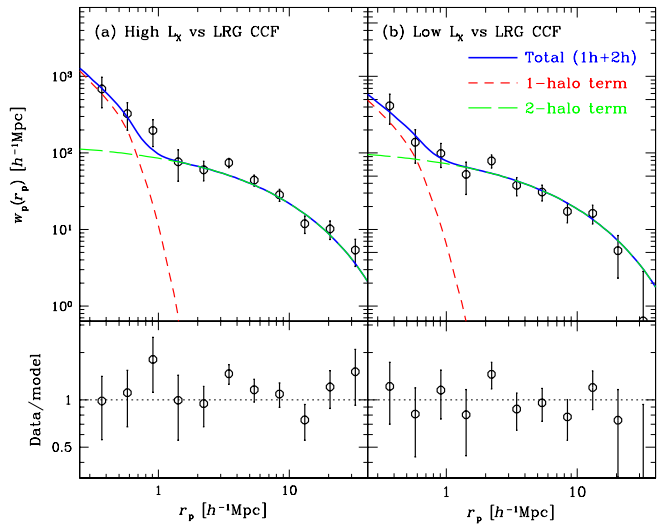}
 \caption{The AGN-LRG CCFs and the best-fit HOD models (Model A) for the (a) high $L_{\rm X}$ and
 (b) low $L_{\rm X}$ RASS-AGN samples. The meanings of line styles and symbols are the 
  same as those in Fig. \ref{fig:tot_ccf_con}(a).
 \label{fig:hxlx_ccf_hod}}
\end{figure}

\begin{figure}
\centering
\includegraphics[width=0.7\hsize]{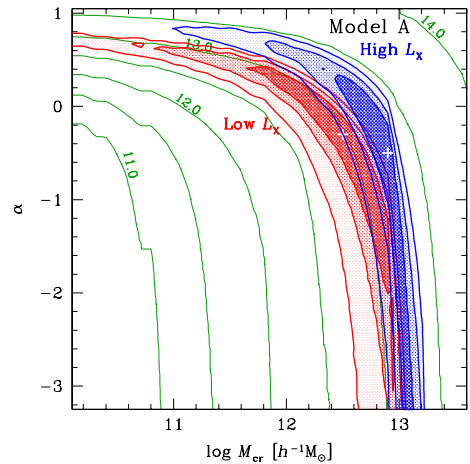}
 \caption{The confidence contours of the fits to the CCFs to Model A
   for the high $L_{\rm X}$ (thick blue lines) and the low $L_{\rm X}$ (thick red lines) RASS-AGN samples. The contours 
   are at $\Delta \chi^2=1.0,2.3$ and $4.6$ levels. The confidence contours are shaded 
   for the ease of the visibility of the difference between the two samples. The 
   $\log\, \langle M_{\rm h}\rangle$ contours are shown in the same way 
   as in Fig. \ref{fig:tot_ccf_con}(b). 
   \label{fig:hxlx_con}}
\end{figure}

Figure \ref{fig:tot_ccf_con}(b) shows the confidence contours in the
$\log\,M_{\rm cr}$ -- $\alpha$ space, overlaid on underlying thin (green) contours 
showing the mean halo mass $\langle M_{\rm h}\rangle$. \chg{\alc{A line of 
constant bias, $b_{\rm A}$, roughly follows a constant $\langle M_{\rm h}\rangle$ 
contour, with small deviations caused by the non-linearity of the $b_{\rm h}(M_{\rm h})$ 
function. The degree of the deviation is such that 
$\log \langle M_{\rm h}\rangle$ for a given $b$ decreases typically by 
$\approx 0.4$ when $\alpha$ is decreased from  $\approx 1$ to $\approx -3$.}} 
Figures \ref{fig:hxlx_ccf_hod} and \ref{fig:hxlx_con} show our HOD fits {and confidence 
contours, respectively,} for the high $L_{\rm X}$ and low $L_{\rm X}$ 
RASS-AGN samples, with an emphasis on the comparison between the two.

 Figures \ref{fig:tot_ccf_con}(b) and \ref{fig:hxlx_con} show that the 
parameter pair $(\log\,M_{\rm cr}, \alpha)$ is  tightly constrained roughly along 
the constant $\langle M_{\rm h}\rangle$ line. This primary
constraint comes from the amplitude of the 2-halo term, which is proportional
to the bias parameter of the AGN sample (for the fixed bias of the LRG sample).
In the HOD analysis, however, the inclusion of the 1-halo term adds additional
constraints in the two-parameter space.

 In any of the total, high $L_{\rm X}$ and low $L_{\rm X}$ RASS-AGN samples, models where 
the number occupation of AGNs is proportional to $M_{\rm h}$ above $M_{\rm cr}$ ($\alpha=1$)  are excluded,  
while a constant number of AGNs per halo above $M_{\rm cr}$ is preferred. 
Figs \ref{fig:tot_ccf_con}(b) and \ref{fig:hxlx_con} put constraints on the values of 
$M_{\rm cr}$ for each sample. Using the $\Delta \chi^2=2.3$ contour (68\% confidence level for 
two parameters), the minimum mass is constrained to be 
$\log\, M_{\rm cr}[h^{-1}{\rm M}_\sun] \ga$ 11.9, 11.9, \& 10.7 for the total, 
high $L_{\rm X}$ and low $L_{\rm X}$ RASS-AGN samples respectively. For the total 
RASS-AGN sample, the contour is constrained to have $\alpha>-3.0$, while for the high $L_{\rm X}$ 
and low $L_{\rm X}$ RASS-AGN samples, the contour allows the smallest $\alpha$ that we have 
explored ($\alpha=-3.25$). This essentially gives constraints on the {\em width} of 
the HOD, and at least for the total RASS-AGN sample, a delta-function type HOD could be 
marginally excluded. 

\begin{deluxetable*}{ccccccc}
\footnotesize
\tablecaption{The HOD parameters for RASS AGNs at $z\approx 0.28$ (Model A)\label{tab:ccf_fit}}
\tablewidth{0pt}
\tablehead{
\colhead{RASS-AGN} &\colhead{$\log M_{\rm cr}$\tablenotemark{a}}&\colhead{$\alpha$\tablenotemark{a}}&
\colhead{$\langle N_{\rm A}\rangle(M_{\rm cr})$\tablenotemark{b}}&
\colhead{$\log\,\langle M_{\rm h}\rangle$\tablenotemark{c}}&\colhead{$b_{\rm A}$\tablenotemark{c}} & \colhead{$\chi^2$\tablenotemark{e}}\\
\colhead{Sample}&\colhead{[$h^{-1} {\rm M}_{\sun}$]}&\colhead{}&\colhead{}&
\colhead{[$h^{-1} {\rm M}_{\sun}$]}&\colhead{}&\colhead{}
}
\startdata
\\
Total            & 12.6 (11.9; 13.0) & -0.3 (0.4;-3.0) & 8$\times 10^{-2}$ & 13.09$\pm$0.08 & 1.30$\pm$0.09 & 6.9 \\  
High $L_{\rm X}$ & 12.9 (11.9; 13.2) & -0.5 (0.7;-3.25\tablenotemark{d}) & 3.3$\times 10^{-3}$&13.26$^{+0.07}_{-0.09}$ 
& 1.44$^{+0.04}_{-0.11}$ & 10.9\\  
Low $L_{\rm X}$  & 12.5 (10.7; 13.0) & -0.3 (0.7;-3.25\tablenotemark{d}) & 5.5$\times 10^{-2}$ & 12.97$^{+0.17}_{-0.13}$
& 1.22$\pm$0.15 & 12.5\\
\enddata
\tablenotetext{a}{The 68\% confidence range in the 2D parameter space ($\Delta \chi^2<2.3$).}
\tablenotetext{b}{The normalization at the nominal case.}
\tablenotetext{c}{The 68\% error ($\Delta \chi^2<1$) for one parameter.}
\tablenotetext{d}{The bound truncated at our parameter grid limit.}
\tablenotetext{e}{The $\chi^2$ value calculated over 11 data points with a covariance matrix.}
\end{deluxetable*}

 The best fit HOD parameters, the mean DMH mass occupied by the AGNs, and the 
linear bias from the fits are summarized in Table \ref{tab:ccf_fit}. For 
the fitting parameters $\log\,M_{\rm cr}$ and $\alpha$, we show the full range 
corresponding to 68\% confidence range  in the 2D parameter space 
($\Delta \chi^2=2.3$).  As shown in Figs. \ref{fig:tot_ccf_con} and 
\ref{fig:hxlx_con}, these two parameters are highly correlated and the confidence contours are 
skewed, thus one should be cautious in interpreting these ranges. On the other hand, if we project 
the probability distribution on the variable $\langle M_{\rm h} \rangle$ or the variable $b$ (bias), 
both of which are unique functions of our two fitting parameters, the projected probability 
distribution becomes roughly Gaussian. Thus we take the $\Delta \chi^2 < 1$ range 
(68\% error for one parameter) 
to estimate the 1$\sigma$ errors of these derived quantities, $\langle M_{\rm h} \rangle$ 
and $b_{\rm A}$, in Table \ref{tab:ccf_fit}.

 \chg{We note that because of the exponential drop of the DMH mass function the HOD at very high $M_{\rm h}$ 
\alc{ contributes little} to the CCF. In order to check the maximum
$M_{\rm h}$ that our analysis can reasonably constrain, we recalculate the CCF with the
truncated power-law model with an upper mass cutoff:
\begin{equation}
\langle N_{\rm A}\rangle \propto M_{\rm h}^\alpha\;\Theta(M_{\rm h}-M_{\rm cr})\Theta(M_{\rm cut}-M_{\rm h})\nonumber
\label{eq:agnhod_cut}
\end{equation} 
 We recalculate $\chi^2$ with varying $M_{\rm cut}$ while fixing ($M_{\rm cr}$,$\alpha$) at
the best-fit value above. The change of $\chi^2$ from the
$M_{\rm cut}=\infty$ is larger than unity at $\log M_{\rm h} [h^{-1}M_\sun]<14.0$. Thus our  
HOD analysis is sensitive up to $M_{\rm h}\approx 10^{14}h^{-1}M_{\rm \sun}$, i.e., the mass of a poor cluster.} 
       

\subsection{HODs and Space Densities}

\begin{figure}
 \centering
 \includegraphics[width=\hsize]{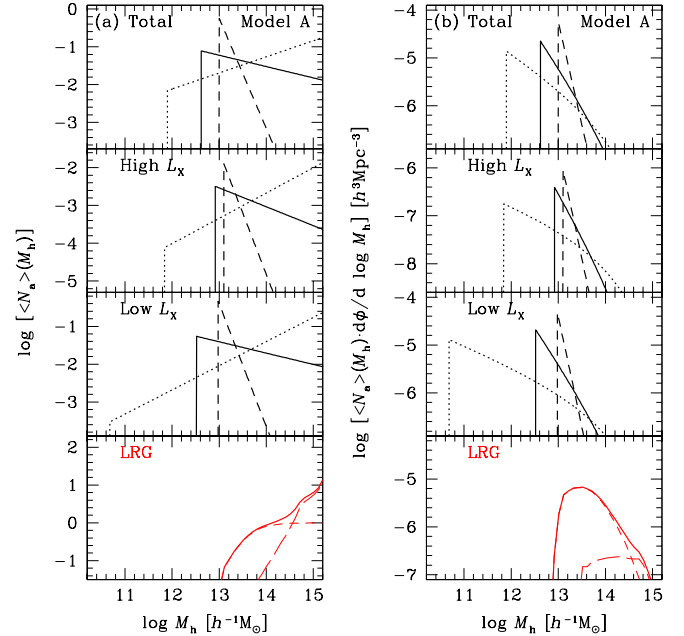}
 \caption{(a) \chga{The HODs ($\langle N_{\rm A}\rangle (M_{\rm h})$) of Model A for three representative 
   points in the parameter space that are acceptable ($\Delta \chi^2<2.3$)} from our fits 
   for each of the total (top panel), high $L_{\rm X}$ (upper middle panel) and low $L_{\rm X}$ 
   (lower middle panel) AGN samples. The solid, 
   dotted, and dashed lines represent the (1) best-fit, (2) lowest $M_{\rm cr}$ extreme, and
   (3) lowest $\alpha$ extreme respectively (see text). \chg{For reference, the bottom panel shows 
   the LRG HODs for the central galaxies (short-dashed line), satellites (long-dashed line), and
   their sum (solid line).}
    (b) The same set of models are shown in terms of the spatial number density 
      per $\log\,M_{\rm h}$. The meaning of the line styles is the same as those in
      panel (a).}
   \label{fig:hods_dens}
\end{figure}

 In order to illustrate the range of acceptable HODs, we plot a number of representative AGN HOD models accepted 
by our fits ($\langle N_{\rm A} \rangle (M_{\rm h})$) as function of DMH mass  
in Fig. \ref{fig:hods_dens}(a). In Fig. \ref{fig:hods_dens}(b), we also show the same sets of models in units 
of the spatial density per comoving volume per log of the DMH mass:
\begin{equation}  
\langle N_{\rm A} \rangle(M_{\rm h}) \frac{d \phi (M_{\rm h})}{d \log\,M_{\rm h}} \equiv
\langle N_{\rm A} \rangle(M_{\rm h}) \cdot \ln(10)M_{\rm h}\phi(M_{\rm h}).
\end{equation}   
These models have been normalized to the observed number densities (see Table \ref{tab:samp}). 
In each panel, three curves are plotted, representing the best-fit case and two extreme cases. 
These are (1) for the best-fit values (solid lines),  (2) the point on the $\Delta \chi^2 = 2.3$ 
contour that has the smallest $M_{\rm cr}$ value (i.e., the upper left tip of the two-parameter 
68\% confidence area in Fig. \ref{fig:tot_ccf_con}(b) or \ref{fig:hxlx_con}) (dotted lines), and 
(3) the smallest $\alpha$ point on the  $\Delta \chi^2 = 2.3$ (the bottom of the confidence region 
and narrowest possible HOD distribution, dashed lines). The best-fit model with $\alpha=-3.25$ 
(the smallest $\alpha$ in our search grid) is used {for the high $L_{\rm X}$ and low $L_{\rm X}$ 
RASS-AGN samples, for which the $\Delta \chi^2 = 2.3$ contour continues below $\alpha=-3.25$.} 
These three models have been plotted to illustrate the almost full 
range of possible HODs statistically accepted by our analysis. 
\chg{In order to illustrate the 
DHM mass range that can be occupied by both LRGs and AGNs, and therefore contributes to
the 1-halo term of the CCF, we also show the HODs of the LRGs for reference.} 
Of course, these plots are for our rather restrictive truncated power-law model and 
by no means are intended to illustrate an exhaustive set of possible HODs. 
However, this illustrates that these three models have roughly the same average 
$M_{\rm h}$, with varied widths of the HOD. An important constraint is that the AGN HOD can 
be no wider than the dotted line in \ref{fig:hods_dens}(a)(b), which represents the 
point corresponding to the lowest extreme of $M_{\rm cr}$ and the roughly highest 
extreme of $\alpha$, among acceptable models with $\Delta \chi^2 < 2.3$.  The meaning of 
this constraint is discussed in Sect. \ref{sec:disc_imp}.


\subsection{Effects of Central AGNs }
\label{sec:modelB}

\chg{In this section we explore models in which 
central AGNs are explicitly included, using galaxy
HODs as a template. Galaxy HODs have been extensively investigated using, e.g.,  SDSS data, 
with a high statistical accuracy. A simple model for luminosity-threshold 
galaxy samples (i.e. more luminous than a given limit), introduced by \citep{zehavi05a} 
is a three parameter model including a step function for the central HOD and a truncated power-law satellite HOD. 
We express the AGN HODs here with the same form, except that the global normalization
is also a free parameter (Model B):  
\begin{eqnarray}
\langle N_{\rm A,c}\rangle &=& f_{\rm A}\Theta(M_{\rm h}-M_{\rm min}), \nonumber \\
\langle N_{\rm A,s}\rangle &=& f_{\rm A}\Theta(M_{\rm h}-M_{\rm min})(M_{\rm h}/M_1)^{\alpha_{\rm s}},  
\label{eq:step_pl}
\end{eqnarray}
where $f_{\rm A}$ is the AGN fraction (duty cycle) among central galaxies at $M_{\rm h}\ga M_{\rm min}$.
We use the symbol $\alpha_{\rm s}$ to emphasize that it represents the HOD slope for satellites. 
$M_1$ is the DMH mass at which the number of central AGNs is equal to that of satellite AGNs.
For luminosity-threshold galaxy samples, it is usually assumed that the central galaxy HOD saturates at unity
at the high $M_{\rm h}$ end (i.e., $f_{\rm A}=1$ in Eq. \ref{eq:step_pl}), because central galaxies in the highest mass 
DMHs are expected to be luminous enough to be included in the sample (e.g. Sect. \ref{sec:lrghod}). 
Since only a small fraction of galaxies contain an AGN, this has to be multiplied by a factor $f_{\rm A}(<<1)$, 
the value of which does not affect the CCF as discussed above. The value of $f_{\rm A}$ can be determined by 
normalizing the HOD to $n_{\rm AGN}$. 

 It is important to note the underlying physical assumption of Eq. \ref{eq:step_pl}, which is that the AGN duty 
cycle among central galaxies does not depend on the DMH mass. We explore the consequences of this model 
as an illustrative contrast to Model A, which assumes a suppression of the AGN duty cycle in central
galaxies of high mass DMHs.} 
  
\chg{We investigate the behavior of $\chi^2$ (Eq. \ref{eq:chi2ccf}) 
in the space of $M_{\rm min}$, $M_1$,
and $\alpha_{\rm s}$. Due to the low signal-to-noise ratio of the CCFs 
for our luminosity-divided  subsamples, we limit our discussion to the total sample. 
Even with the total sample, our statistics do not allow us to constrain 
all three parameters; therefore we investigate the constraints on two parameters by fixing the 
remaining parameter. The HOD analysis of luminosity-threshold samples with $-21.5 \le M_r^{\rm max}\le -21.0$, 
studied by \citet{zehavi05a} found $M_1/M_{\rm min} \approx 23$ and $\alpha_{\rm s}\approx 1.2$.} 
\chg{We consider the typical HOD of SDSS galaxy samples in this luminosity threshold range as a template 
because their $M_{\rm min}$ range ($12.0\la \log M_{\rm min}[h^{-1}M_\sun]\la 12.7$) roughly 
coincides  with the minimum DMH range for our AGNs, as we see from the range of $M_{\rm cr}$ in the 
confidence contour in Fig. \ref{fig:tot_ccf_con}(b), which approximately coincide with 
the range of $M_{\rm min}$ as we see below.  We therefore (i) fix $M_1/M_{\rm min}=23$ and obtain 
constraints in the $(\log\,M_{\rm min},\alpha_{\rm s})$-space, and (ii) fix $\alpha=1.2$ and 
obtain constraints in the $(\log\,M_{\rm min},M_1/M_{\rm min})$-space.}

\begin{figure}
 \begin{center}
   \includegraphics[width=\hsize]{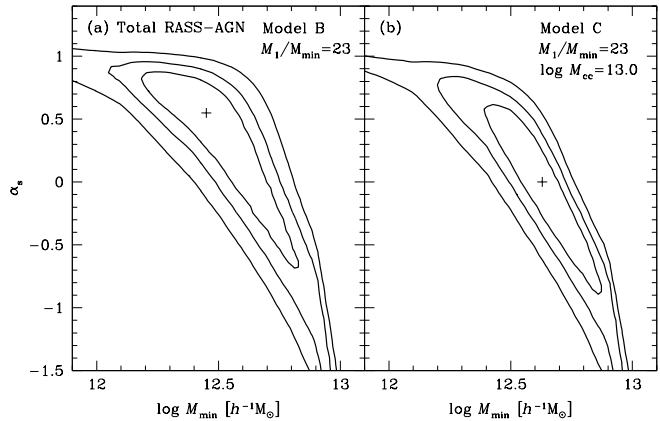}
 \end{center}

 \caption{(a) The confidence contours of the Model B parameters in the ($\log\,\,M_{\rm min}$,$\alpha_{\rm s}$)
   space for $M_1/M_{\rm min}=23$. (b) The confidence contours of the same set of parameters
   for Model C, for $\log M_{\rm cc}[h^{-1}M_\sun]=13.0$ and $M_1/M_{\rm min}=23$.
   The contours are at $\Delta \chi^2=1.0,2.3$ and $4.6$ levels.
   \label{fig:modelBCcont}
 }   
\end{figure}

\begin{deluxetable*}{cccccccc}
\footnotesize
\tablecaption{The HOD parameters for the RASS AGNs (Models B \& C)\label{tab:modelBC_fit}}
\tablewidth{0pt}
\tablehead{
\colhead{Model}&
\colhead{RASS-AGN} &\colhead{$\log M_{\rm min}$\tablenotemark{a}} & \colhead{$M_{\rm cc}$} & 
\colhead{$\log M_1/M_{\rm min}$\tablenotemark{a}} &
\colhead{$\alpha_{\rm s}$\tablenotemark{a}}&\colhead{$f_{\rm A}$\tablenotemark{c}} & \colhead{$\chi^2$\tablenotemark{d}}\\
\colhead{}&\colhead{Sample}&\colhead{[$h^{-1} {\rm M}_{\sun}$]} & \colhead{[$h^{-1} {\rm M}_{\sun}$]}&\colhead{}&\colhead{}
&\colhead{} &\colhead{}
}

\startdata
\\
B  &Total& 12.5 (12.2;13.0) & ... & 1.36 (fixed) & 0.55 (0.95;-1.50\tablenotemark{b}) & 2.9$\times 10^{-2}$  & 6.9\\
B &Total& 12.4 (11.9;12.6) & ... &  1.65 (1.51;1.83) & 1.20 (fixed) & 3.3$\times 10^{-2}$  &  8.4 \\
B &Total& 12.3 (12.1;12.7) & ... & 1.55 (1.36;1.81) & 1.00 (fixed) & 2.6$\times 10^{-2}$  &  7.7 \\
C  &Total& 12.6 (12.2;13.0\tablenotemark{b}) & 13.0 (fixed) & 1.36 (fixed) & 0.0 (0.84;-1.50\tablenotemark{b}) & 
            3.7$\times 10^{-2}$ & 6.8\\
\enddata
\tablenotetext{a}{The 68\% confidence range in the 2D parameter space ($\Delta \chi^2<2.3$).}
\tablenotetext{b}{The bound truncated at our parameter grid limit.}
\tablenotetext{c}{The AGN fraction of the central galaxy evaluated at the best-fit parameters.}
\tablenotetext{d}{The $\chi^2$ value calculated over 11 data points with a covariance matrix.}
\end{deluxetable*}

\chg{ The resulting confidence contours for (i) are shown in Fig. \ref{fig:modelBCcont}(a).  
The best-fit parameters and $\chi^2$ values are summarized in Table \ref{tab:modelBC_fit} for both
(i) and (ii), where we also show the $\alpha_{\rm s}=1.0$ case for (ii), in which the number of satellites 
is proportional to the DMH mass. 

For the $M_1/M_{\rm min}=23$ case, the best-fit slope for the satellite is 
below unity ($\alpha_{\rm s}=0.55$) and $\chi^2=6.9$ is identical to that of Model A. 
The best-fit model with $\alpha_{\rm s}=1.0$ is marginally rejected ($\Delta \chi^2=3.6$) 
and the model with $\alpha_{\rm s}=1.2$ is significantly rejected ($\Delta \chi^2=18$). Models 
that are consistent with $\alpha_{\rm s}=1.2$ can be found at $M_1/M_{\rm min}\geq 30$, but they
are not consistent with the HOD of luminosity threshold galaxy samples in the 
$-21.5 \le M_r^{\rm max}\le -21.0$ range.}


\chg{\chga{We also explore a model (Model C) where only lower mass DMHs contain a central AGN, where the 
physical picture is explained at the introduction of Model A (Sect. \ref{sec:agnhod_ex}). In this model, 
the central HOD in Model B is replaced by
\begin{equation}
\langle N_{\rm A,c}\rangle = f_{\rm A}\Theta(M_{\rm h}-M_{\rm min})\Theta(M_{\rm cc}-M_{\rm h}), 
\label{eq:modelC}
\end{equation}
while the form of $\langle N_{\rm A,c}\rangle$ stays the same. As a representative 
case, we choose $\log M_{\rm cc}[h^{-1}M_\sun]=13.0$, below which the DMH center 
is more frequently occupied by blue galaxies than red galaxies in the HOD analysis of 
color-separated samples by \citet{zehavi05a}, while we keep $M_1/M_{\rm min}= 23$
(but see also \citet{zehavi10}, which use a different parametrization on color-separated samples).
The confidence contours are shown in Fig. \ref{fig:modelBCcont}(b) and 
the  best-fit parameters and $\chi^2$ values are shown in Table \ref{tab:modelBC_fit}. In this scenario, the best-fit slope for 
the satellite is $\alpha_{\rm s}=0$, while $\alpha_{\rm s}=1.0$ is rejected at the $\Delta \chi^2=4.8$ 
level.}}  


\section{Discussion}
\label{sec:disc}

\subsection{Comparison with Results from Paper I}
\label{sec:comp_paperI}


\begin{deluxetable*}{cccccccc}
\small
\tablecolumns{8}
\tablecaption{Comparion of Results from Paper I\label{tab:paperIcomp}}
\tablewidth{0pt}
\tablehead{
\colhead{(1)} &\colhead{(2)}&  \colhead{(3)} &
\colhead{(4)} & \colhead{(5)} & \colhead{(6)} & \colhead{(7)} & \colhead{(8)}\\ 
\colhead{RASS-AGN} &\colhead{$b_{\rm A}$}&  \colhead{$b_{\rm A}$} & \colhead{$b_{\rm A}$} 
& \colhead{$\log \langle M_{\rm h}\rangle$} 
& \colhead{$\log (M_{\rm h,typ})$\tablenotemark{a}} & \colhead{$\log (M_{\rm h,typ})$\tablenotemark{a}} 
& \colhead{$\log (M_{\rm h,typ})$\tablenotemark{a}}
\\
\colhead{Sample} & \colhead{HOD} & \colhead{PL} & \colhead{PL,$>1.5$\tablenotemark{b}} &
\colhead{HOD,tin05\tablenotemark{c}} & \colhead{HOD,tin05\tablenotemark{c}} & 
\colhead{PL,smt01\tablenotemark{c}} & \colhead{PL,tin05\tablenotemark{c}} \\
\cline{5-8}
\colhead{} & \colhead{} & \colhead{} & \colhead{} 
& \multicolumn{4}{c}{[$h^{-1} {\rm M}_{\sun}$]}}
\startdata
\\
Total           &1.32$\pm$0.08         &1.11$^{+0.10}_{-0.12}$ & 1.52$\pm 0.21$ & 
                13.09$\pm$0.08 & 13.00$\pm$0.12 
                &12.58$^{+0.20}_{-0.33}$&12.70$^{+0.19}_{-0.30}$\\  
High $L_{\rm X}$&1.44$^{+0.04}_{-0.11}$&1.44$^{+0.22}_{-0.27}$ & 1.56 $\pm 0.23$ &
                13.26$^{+0.07}_{-0.09}$& 13.19$^{+0.05}_{-0.12}$
                &13.10$^{+0.24}_{-0.43}$&13.27$^{+0.22}_{-0.41}$\\  
Low $L_{\rm X}$ &1.23$\pm$0.15         &0.88$^{+0.14}_{-0.17}$& 1.6$^{+2.1}_{-0.6}$ &
               12.97$^{+0.17}_{-0.13}$& 12.87$^{+0.13}_{-0.30}$
                &11.83$^{+0.55}_{-\infty}$&12.03$^{+0.39}_{-1.48}$\\ 
\enddata
\tablenotetext{a}{The symbol $M_{\rm typ}$ represent the DMH mass satisfying $b_{\rm h}(M_{\rm typ})=b_{\rm A}$.}
\tablenotetext{b}{The range of $r_{\rm p}$[$h^{-1}$Mpc] used in the power-law limited to 
$1.5<r_{\rm p}[h^{-1}{\rm Mpc}]<15.$}
\tablenotetext{c}{Conversion from bias to $M_{\rm h}$: smt01 \citep{sheth01} or tin05 \citep{tinker05}.}
\end{deluxetable*}

 Table \ref{tab:paperIcomp} compares our results with those from paper I. 
The meanings of the table columns are: (1) sample, (2) the bias parameter from the HOD analysis 
(Eq. \ref{eq:bias_A}), 
(3) the bias parameter from paper I, i.e. for the power-law fit in 
$0.3<r_{\rm p}[h^{-1}{\rm Mpc}]<15$,  (4) the bias parameter for the power-law fit in 
$1.5<r_{\rm p}[h^{-1}{\rm Mpc}]<15$ calculated in the same way as paper I (see below),
(5) the ``mean'' DMH mass from Eq. \ref{eq:meanmass}, 
(6) the ``typical'' DMH mass calculated from $b_{\rm A}$ from the HOD model, \chg{defined
by $b_{\rm h}(M_{\rm typ})=b_{\rm A}$} calculated with the $b_{\rm h}(M_{\rm h})$ 
by \citet{tinker05} (see Sect. \ref{sec:model_ing}).
(7) the ``typical'' DMH mass from paper I \chg{(i.e. from power-law fits)} calculated 
with the $b_{\rm h}(M_{\rm h})$ by \citet{sheth01}, and (8) the ``typical'' DMH mass 
from the paper I bias parameter
re-calculated using the $b_{\rm h}(M_{\rm h})$ relation by \citet{tinker05}. While the results 
from the the HOD analysis presented here and the simple analysis of paper I agree well 
for the high $L_{\rm X}$ RASS-AGN sample, the HOD analysis gives larger bias parameters than 
those derived in paper I for the total and low $L_{\rm X}$ RASS-AGN samples at the level of 
1.5 $\sigma$. Since these bias parameters have been derived from the same dataset with different 
methods, the discrepancies are systematic rather than statistical. 
One of the major findings of paper I is the detection of the X-ray luminosity
dependence of the correlation function. The difference in correlation lengths ($r_0$),
between the high $L_{\rm X}$ and low $L_{\rm X}$ RASS-AGN samples measured in paper I is 
$\approx 2.5\sigma$, based on power-law fits with a fixed slope of $\gamma=1.9$. 
The differences of the AGN bias parameters $b_{\rm A}$ and associated typical DMH mass
$M_{\rm h,typ}$ derived from power-law fits are $\approx 1.8\sigma$. In the HOD analysis, 
the significance of the difference in $b_{\rm A}$ decreases to  $1.1\sigma$, because it is 
\chg{constrained mainly by measurements at large scales where the 2-halo term dominates (see below).} 
However, in terms of $M_{\rm h,typ}$ derived 
from the HOD analysis, the difference is $\approx 1.8\sigma$ because of the additional 
constraints from the 1-halo term. 
 
 There are a number of differences in the processes used to derive the bias parameter between 
paper I and this work. In paper I, the bias parameters are based on the power-law fits to 
Eq. \ref{eq:acf_infer} over scales of  $0.3<r_{\rm p} [h^{-1}{\rm Mpc}] <15$, assuming a linear bias 
over the entire scale range. Then the power-law models are converted into the rms fluctuations over 
8 $h^{-1}$ Mpc spheres  ($\sigma_{\rm 8,AGN}$), and the bias parameter is calculated using 
$\sigma_{\rm 8,AGN}/[\sigma_0D(z)]$, where $D(z)$ is the linear growth factor.
 The main source of discrepancy is that, while paper I also uses data points down to 
a scale of $r_{\rm p}=0.3 h^{-1}$ Mpc to obtain $b$, in the HOD analysis presented here 
the constraints on $b$ mainly come from the 2-halo term dominated 
region ($r_{\rm p}\ga 1.5 h^{-1}$), while smaller scale data constrain other variables. 
Fig. \ref{fig:hxlx_ccf_hod} shows that the difference between the high $L_{\rm X}$ 
and low $L_{\rm X}$ RASS-AGN samples is more prominent in the 1-halo term dominated 
regime, at $r_{\rm p}\la 1 h^{-1}$, than in the 2-halo term dominated regime.
{This means that, for the AGN-LRG CCF, the 1-halo term is more sensitive
to differences in the AGN HOD. The reason for this is that while $b_{\rm h}(M_{\rm h})$ 
increases only weakly with increasing with $M_{\rm h}$, the 1-halo term is directly proportional 
to the number of AGNs that are in the same DMHs as those occupied by LRGs, 
i.e., very massive ones ($\log M_{\rm h}\ga 13.5 [h^{-1}{\rm M_{\sun}}]$), \chg{i.e.
the 1-halo term is more sensitive to the existence of AGNs in high mass halos.}} 

 In the literature, relative or absolute bias parameters are often calculated based 
on results from power-law fits \citep[e.g., paper I;][]{mullis04,yang06,miyaji07,coil09}. 
The power-law models of $\xi(r)$ are usually converted to the rms fluctuation over 
8$h^{-1}$ Mpc spheres or are averaged up to the distance of 20$h^{-1}$ Mpc 
($\bar{\xi}(<20h^{-1}{\rm Mpc})$).  While some authors use only large scales ($r_{\rm p}> 1-2 h^{-1}$Mpc) 
to ensure that only the linear regime is used, others include smaller scales. Usually the rationale 
for including smaller scales is to have better statistics, combined with the 
empirical observation that the correlation functions 
are well fit with a power-law model over the scale of 
$0.1-0.3\la r_{\rm p}\la 10-20 h^{-1}$Mpc ($0.3<r_{\rm p}<15 h^{-1}$ Mpc in the case of paper I). 
As a check, we have re-calculated the bias parameter for power-law fits to the 
$w_p(AGN|AGN)$ used in Paper I on scales of  $1.5< r_{\rm p}<15 h^{-1}$ Mpc, to limit ourselves to
the linear regime. These are shown in 
column (4) of Table \ref{tab:paperIcomp}. The bias parameters obtained in this manner
carry large statistical errors, but the values are consistent with those derived from the HOD analysis 
within 1$\sigma$ in all three samples, while they are more than $\approx 1 \sigma$ above the 
bias parameters derived in paper I for the total and low $L_{\rm X}$ RASS-AGN samples.
\chg{The AGN bias parameters in our total and high $L_{\rm x}$ samples from the HOD analysis, 
as well as the power-law fit $1.5< r_{\rm p}<15 h^{-1}$ Mpc, 
are consistent within combined 1$\sigma$ errors with those measured by \citet{padmanabhan09} for 
their QSO samples ALL0 and LSTAR0, which covers a similar redshift range as ours ($0.25<z<0.35$).
However, their bias parameters for their higher redshift QSO samples, ALL1 \& LSTAR1 ($0.33<z<0.50$) 
as well as ALL2 \& LSTAR2 ($0.45<z<0.60$) are lower.}

 Using the power-law fit results including the non-linear regime is useful to detect 
the different clustering properties between, e.g., different source populations, luminosities,
or types. In our case, this gives the most statistically significant difference
between the clustering properties of the high $L_{\rm X}$ and low $L_{\rm X}$ RASS-AGN samples. 
 However, one has to be careful in interpreting the results, e.g., in terms of the DMH 
mass based on the bias parameter. In principle, the HOD analysis overcomes this limitation, 
because we are able to  derive the pure linear bias parameter, while at the same time 
utilizing the data in the non-linear regime.     
   
\subsection{Implications for the Environments of AGN Accretion}
\label{sec:disc_imp}

 With the HOD analysis, we can also obtain constraints on at least one additional 
variable which is independent from the ``mean'' DMH mass, thanks to the 
constraints from the 1-halo term. \chg{\alc{The importance of the 1-halo term
constraints has also been noted by \citet{padmanabhan09} using a similar 
CCF analysis between photometric-redshift LRG samples and optically-selected
QSOs in SDSS at $z\sim 0.5$. They qualitatively concluded that at least some
QSOs must be satellites and also that models of the form of our Model B  
(Eq. \ref{eq:step_pl}) with $M_1/M_{\rm min}=20$ and $\alpha_{\rm s}=0.9$ 
 are also acceptable, while lower $\alpha$ values are also possible. Our analysis here 
 gives qualitatively similar results.} }

 \chg{The acceptable parameter space for Model A is always in the $\alpha<1$ 
regime for all of our three AGN samples: the total, high $L_{\rm X}$, and low $L_{\rm X}$ 
RASS-AGN samples.}  This means that the average number of AGNs in a DMH does not increase 
proportionally with mass. Figs.~\ref{fig:tot_ccf_con}(b) and \ref{fig:hxlx_con}
show that models with $\alpha\la 0$ are preferred. 
It is interesting to compare this result with HOD analyses of galaxies,
that find $\alpha_{\rm s}\approx 1-1.2$ for a wide range  of absolute magnitudes and 
redshifts at least up to $z\approx 1.2$ \citep{zehavi05a,zheng07,zehavi10}.
In principle, neither the results nor the assumptions of Model A say anything 
about the central fraction among AGNs at the lower mass DMHs  (see Sect. \ref{sec:agnhod_ex}), 
and thus they do not directly constrain the slope of the satellite AGN HOD $\alpha_{\rm s}$.  
However, if we take the extreme case where all the AGNs are satellites, $\alpha_{\rm s}<0.4$, 
this would imply that AGN fraction among satellite galaxies strongly decrease with $M_{\rm h}$. 
\chg{\chga{In Model B, where the central galaxy component is included explicitly under the assumption
that the AGN duty cycle in central galaxies is constant against varying $M_{\rm h}$
above $M_{\rm min}$, the best-fit slope is still less than unity ($\alpha_{\rm s}=0.55$), 
and models with $\alpha_{\rm s}\approx 1$ are only marginally rejected for 
$M_1/M_{\rm min}=23$ (a representative value of luminosity-threshold galaxy samples). 
In case of Model C, where only  $\log\,M_{\rm h}[h^{-1}M_\sun]<13.0$ DMHs can contain a central 
AGN, the best-fit solution is $\alpha_{\rm s}=0$, while $\alpha_{\rm s}\approx 1$ is rejected
at a $\sim 2\sigma$ level for $M_1/M_{\rm min}=23$.  On the whole, our results tend to favor models in 
which the AGN fraction among satellite galaxies {\it decreases} with increasing DMH mass 
(or the richness of groups and clusters), in the mass range 
$12-13\la \log\, M_{\rm h} [h^{-1}{\rm M_\sun}]\la 14.0$.}}
  
 Since our AGN sample is selected from X-ray point sources in RASS, and since clusters of galaxies 
(which represent the most massive DMHs) are also X-ray sources, an \chg{observational bias that
might affect our results is that AGNs in the clusters are {\em selected against} in the 
RASS-AGN sample} due to confusion, {especially considering that the mean point spread function (PSF)
has a FWHM$\approx 2$ arcminutes \citep{labarbera09}, which is not negligible 
when compared to the extent of the cluster X-ray emission.} We have estimated the significance of 
this effect as follows. The detection limit of the cluster diffuse X-ray emission is higher than 
that of point sources. A typical flux limit for clusters in RASS is 
$S_{\rm X}\approx 2\times 10^{-12}$ erg s$^{-1}$ cm$^{-2}$,
(0.1-2.4 keV)\citep{boehringer01,popesso04}, corresponding to $\log\, L_{\rm X}\ga 44.3$ 
$[h_{70}^{-2}{\rm erg\,s^{-1}}]$ at the lower bound of our redshift range of $z\approx 0.16$.  This scales
to a cluster virial mass of $\log\, M_{\rm h}\ga 14.5\, h^{-1}{\rm M_\sun}$ \citep{reiprich02}.
\chg{As we found at the end of Section \ref{sec:results_fit}, our CCF is only sensitive to the HOD 
behavior at $\log\, M_{\rm h}\la 14.0\, h^{-1}{\rm M_\sun}$, and thus this selection effect 
should play negligible role in our results.}
Additionally, a confusion due to two or more X-ray AGN point sources 
being artifically blended into a single source 
by the RASS point spread function (PSF) is not important either, 
as $\langle N_{\rm A}\rangle$ is much less than unity over all masses 
(see Fig. \ref{fig:hods_dens}).  Thus we can state with a good degree of confidence
that our limits on $\alpha$ (Model A) or $\alpha_{\rm s}$ 
(Model B/C) are not an artifact of X-ray source confusion.    
  
 \chg{Our results of $\alpha_{\rm s}<1$, found in our extensive explorations in the 
parameter space of the models that we consider, may be interpreted in view of the 
long-suggested deficiency of emission line diagnostic-selected AGNs in rich 
clusters \citep[e.g.][]{gisler78,dressler85}. While our CCF analysis is not sensitive to rich 
clusters (Sect. \ref{sec:results_fit}), a more recent work by \citet{popesso06} verified that this 
trend extends to the group scale by finding an anti-correlation between AGN fraction and the galaxy 
velocity dispersion in nearby ($z<0.08$) groups/clusters of galaxies down to the velocity dispersion 
of $\sigma_{\rm v}\sim 200\,[{\rm km\,s^{-1}}]$.} \citet{arnold09} also found that the X-ray 
selected AGN ($\log\, L_{\rm X}\ga 41 h_{70}^{-1}$erg s$^{-1}$, 
representing a much lower luminosity population than our AGNs) fraction is larger in 
groups than in clusters by a factor of two at low redshifts ($0.02<z<0.06$).
Various physical mechanisms have been suggested to explain the deficiency of
star-formation galaxies in clusters of galaxies, and similar processes might
be responsible for the lack of AGNs as well. These include ram-pressure stripping 
\citep{gunn_gott72} or evaporation \citep{cowie_songaila77} of a galaxy's 
interstellar medium, which is a required ingredient of the AGN activity
in galaxies, by the hot intracluster medium. Also, if AGN activity is triggered by mergers 
between two gas-rich galaxies, another possible mechanism for this trend is the decreased 
cross-section for galaxy mergers in the galaxy-galaxy close encounters at high relative 
velocities \citep{makino97}. In high mass DMHs, representing richer groups/clusters, the velocity 
dispersion is higher and thus one may expect that there are fewer mergers than in groups of galaxies 
\citep{popesso06}, although the density of galaxies increase at the center. Other mechanisms such as 
the accretion of small gas-rich dwarf galaxies (minor mergers) may play a role in fueling the AGN host, but
it is not clear how this would affect the slope $\alpha_{\rm s}$.

In order to fully understand the growth history of SMBHs as well as the physical processes
responsible for the AGN activity in groups and clusters, we need to explore the statistics at 
different redshifts, luminosities, and AGN types. The results mentioned above that investigate AGNs 
in groups/clusters are concerned with AGN at lower redshifts and with lower
luminosities than the AGNs in our CCF analysis. In the AEGIS field,
\citet{georgakakis08} found that AGNs at $z\approx 1$ are more frequently found
in the group environment than the overall galaxy population. However, 
deep field surveys such as AEGIS do not contain a large enough volume to 
explore the richness dependence of the AGN fraction in groups/clusters. 
\citet{martini09} investigate 
Chandra images of a sample of 33 clusters at $0.05<z<1.3$, finding that the luminous  AGN 
($\log\, L_{\rm X}\ga 43 h_{70}^{-1}$erg s$^{-1}$) fraction increases with 
redshift. {They explore the redshift dependence of the AGN fraction, but 
not the richness dependence. At $z<0.4$, where the luminosity range of their AGNs 
is closer to ours, they have only two AGNs  in 17 clusters in the redshift
range that is similar to our sample. A much larger sample of groups/clusters 
is needed to explore the richness dependence of the AGN fraction in a sufficiently 
narrow redshift range to avoid any degeneracy with the redshift dependence.} 

 In general, directly investigating AGN prevalence in individual 
groups and clusters is observationally and analytically intensive. 
The HOD analysis of galaxy-AGN CCF provides a strong statistical tool 
with which to probe the AGN population in groups and clusters, which can be 
made without identifying individual groups and clusters. Thus it is important to extend the HOD analysis to 
AGN clustering measurements at higher redshift, especially in view that the fraction of star-forming
galaxies in clusters increase with redshift \citep{butcher84}. 
\chg{\alc{It is also important to have larger numbers of AGNs for the CCF measurements, which will 
enable measurements of small-scale clustering and produce stronger constraints 
on $\alpha_{\rm s}$, as well as to break the model degeneracy. Using non-LRG galaxies for the CCF 
to obtain the 1-halo term constraints in the lower $M_{\rm h}$ regime would also improve 
the analysis.}}   
   
 Linked to our constraints on $\alpha$ in \chg{Model A}, we also have weak constraints on
the range of $M_{\rm cr}$.  \chg{The range of $M_{\rm cr}$ is similar to that of $M_{\rm min}$
in Model B.} Using the scaling relations and the luminosity range of our 
RASS-AGN samples, we can make an interesting comparison. Figure 1 of paper I shows that the 
minimum 0.1-2.4 keV luminosities of the total and high $L_{\rm X}$ RASS-AGN samples are 
$\log\,L_{\rm X}\,[h_{70}^{-2}{\rm erg\,s^{-1}}] \approx 43.7$ \& 44.3 respectively, 
while the low $L_{\rm X}$ sample has the same minimum luminosity as the total RASS-AGN 
sample. Converting the 0.1-2.4 keV luminosity to a 0.5-2 keV luminosity, 
assuming a power-law spectrum with photon index $\Gamma=2.5$ and applying the 
luminosity-dependent bolometric correction from Eq. (2) of \citet{hopkins07}, 
the corresponding bolometric luminosities are 
$\log\,L_{\rm bol}\, [h_{70}^{-2}{\rm erg\,s^{-1}}] \approx$ 44.5 and 45.4  
for the total and high $L_{\rm X}$ RASS-AGN samples. The mass of the SMBH ($M_{\bullet}$) 
corresponding to the minimum luminosity is then $M_{\bullet} \approx 2\times 10^{6}/\lambda$
and $2\times 10^{7}/\lambda$ $[{\rm M_\sun}]$, where $\lambda \equiv L_{\rm bol}/L_{\rm Edd}$
and $L_{\rm Edd}$ is the Eddington Luminosity. 

 The relation between the black hole mass $M_{\bullet}$ and the DMH of its host galaxy 
is explored by \citet{ferrarese02} by comparing the rotation curves of spiral galaxies. 
While their 
$M_{\bullet}$-DMH mass relation is subject to uncertainties due to different 
assumptions of 
the circular velocity and virial velocity by a factor of a few, using their Eq. (6) with
\citet{bullock01}'s recipe, the above black hole mass corresponds to a galaxy halo mass
of $4\times 10^{11}$ ($1.5\times 10^{12}$) $\cdot \lambda^{-0.6}[{\rm M_\sun}]$ 
(for $h=0.7$). These masses are comparable or slightly lower than the lower bounds 
of $M_{\rm cr}$ derived above for those AGNs emitting at the Eddington luminosity and 
are an order of magnitude lower than our best-fit values. In general, the concept of the 
DMH of a galaxy explored by \citet{ferrarese02} 
is different from the DMH derived in our CCF/HOD analysis,  
since the large-scale correlation function reflects the DMH mass as the largest virialized 
structure the object belongs to. If the largest virialized structure 
represents a group or cluster of galaxies, then there are multiple objects in a DMH, 
and the DMH that is related to \citet{ferrarese02}'s scaling relation represents 
a sub-halo in a larger mass halo. While our statistics do not allow for 
strong constraints,
if $M_{\rm cr}$ is close to our nominal value, then the lowest luminosity AGNs among our
samples should be accreting at a sub-Eddington rate ($\lambda <0.1$) or reside 
in groups.

\section{Conclusion}
\label{sec:concl}

 We have applied a  halo occupation distribution (HOD) model to the 
two-point cross-correlation function (CCF) between broad-line RASS-AGN and luminous red 
galaxies (LRGs) at $0.16<z<0.36$ measured in paper I. We have developed a method of applying the 
HOD model directly to the CCF, where we have used the well-determined HOD of the LRG as a 
reference to constrain the HOD of the AGNs. Our findings are summarized as follows:

\begin{enumerate}
\item From our  HOD analysis of the AGN-LRG CCF, we have found constraints on the HOD 
  of AGNs. We have modeled the AGN HOD as a truncated power-law model \chg{(Model A)} with two parameters: 
  the critical dark matter halo (DMH) mass ($M_{\rm cr}$) below which the AGN HOD is zero 
  (i.e, there are no AGN in halos below this mass threshold), and the slope $\alpha$, where
  the HOD $\propto M_{\rm h}^\alpha$ for $M_{\rm h}\ge M_{\rm cr}$. We have obtained 
  joint constraints in this two-parameter space, $M_{\rm cr}-\alpha$ under the scenario that there is no
  central AGNs in high mass halos.

\item The constraints in this $M_{\rm cr}-\alpha$ space lie mainly along lines of 
   constant large-scale bias and constant mean DMH mass. These provide a more accurate 
   determination of the mean mass of the DMHs in which AGNs reside than the analysis 
   based on power-law fits provided in Paper I. The mean DMH mass we derive is 
   $\langle M_{\rm h} \rangle [h^{-1}{\rm M_\sun}]$
   $=13.09\pm 0.08$, 13.26$^{+0.07}_{-0.09}$, and  12.97$^{+0.17}_{-0.13}$ for the total, 
   high $L_{\rm X}$ and low $L_{\rm X}$ RASS-AGN samples respectively. The 2-halo term 
   of the CCF primarily constrains the mean DMH mass.

\item In all three of our RASS-AGN samples, we have detected a significant contribution 
  of the 1-halo term in the AGN-LRG CCF, indicating that some AGNs are in the same 
  DMHs occupied by LRGs. This gives additional constraints on the AGN HOD beyond the 
  mean DMH mass.

\item The range of acceptable HODs of high $L_{\rm X}$ and low $L_{\rm X}$ RASS-AGN samples 
   are significantly different in the $M_{\rm cr}-\alpha$ space, confirming the paper I's
   results on the X-ray luminosity dependence of AGN clustering properties at the 
   $\approx 2\sigma$ level.

\item The most important constraint we have obtained \chg{using Model A} is an upper limit 
  (corresponding to $\Delta \chi^2=2.3$), of $\alpha\la 0.4$  for the total AGN sample, 
  rejecting $\alpha = 1$. \chg{In the extreme case where all AGNs are satellites, 
  the results show a} sharp contrast with previously derived HODs of galaxies, 
  which show $\alpha\approx 1-1.2$ (i.e., the number of \chg{satellite} galaxies approximately 
  $\propto M_{\rm h}$), across a range of galaxy luminosity and redshift. Taken at face 
  value, this implies that {\em the AGN fraction among satellite galaxies decreases with 
  increasing $M_{\rm h}$.} 
  This is consistent with previous observations that the AGN fraction is smaller in clusters 
  than in groups in the nearby Universe. Possible explanations include the ram-pressure 
  stripping and/or evaporation of galactic gas by the hot intracluster medium and a decreased 
  cross-section for galaxy merging in the high velocity dispersion environment of richer 
  groups/clusters.

\item \chg{We also investigated a model (Model B) which is 
    composed of central and satellite AGNs, where the central HOD is 
    constant ($=f_{\rm A}$) and the satellite HOD has a power-law form 
    $=f_{\rm A} (M/M_{\rm 1})^{\alpha_{\rm s}}$, both at masses above $M_{\rm min}$. This model 
    is based on the results of HOD analyses of galaxies. For $M_1/M_{\rm min}=23$, 
    which is appropriate for galaxies, we obtain $\alpha_{\rm s}\la 0.95$, finding 
    a marginal preference for a picture in which the AGN fraction among satellite galaxies 
    decrease with DMH mass. \chga{With another model (Model C), which has the same form
    as Model B except that only $\log M_{\rm h}[h^{-1}M_\sun]<13.0$ DMHs are 
    allowed to contain a central AGN, we obtain slightly stronger constraints 
    ($\alpha_{\rm s}\la 0.84$) that give preference to this picture.}}

\item We have also obtained a lower limit (corresponding to $\Delta \chi^2=2.3$) of
  $\log\,M_{\rm cr}[h^{-1}{\rm M_\sun}]\la 11.9$ for the total AGN RASS-sample. 
  If the lowest luminosity AGNs in our sample are emitting at the Eddington luminosity, 
  the black hole mass can be scaled to the DMH mass of the galaxy that is comparable to 
  this lower limit of $M_{\rm cr}$. {If the critical DMH mass is at the best-fit value 
  ($\log\,M_{\rm cr}\,[h^{-1}{\rm M_\sun}]\approx 12.6$),  then the lowest $L_{\rm X}$ 
  AGNs in our sample contain SMBHs that accrete at low Eddington ratios and/or 
  reside in group environments.}   
\end{enumerate}


\acknowledgments

We thank John Peacock for providing us with his HOD modeling code. We also thank 
Francesco Shankar for stimulating discussions and the referee for extensive 
comments and suggestions, which have improved the paper very significantly. 
This work has been supported by CONACyT Grant 83564 and DGAPA-Universidad Nacional 
Aut\'onoma de M\'exico (UNAM) Grants 
PAPIIT IN110209 and IN109710 to IA-UNAM-Ensenada as well as the NASA ADP grant 
NNX07AT02G to UCSD.  This research made use of data from the {\sl ROSAT} satellite, 
which was supported by the Bundesministerium f{\"u}r Bildung und Forschung (BMBF/DLR) and 
the Max-Planck-Gesellschaft (MPG), as well as the  Sloan Digital Sky Survey (SDSS), the 
funding of which has been provided by the Alfred P. Sloan Foundation, the Participating 
Institutions, the National Aeronautics and Space Administration, the National Science 
Foundation, the U.S. Department of Energy, the Japanese Monbukagakusho, and the Max Planck Society. 
The SDSS Web site is http://www.sdss.org/. This research also made use of computing facility 
available from Departmento de Superc\'omputo, DGSCA, UNAM.

\end{document}